\documentclass[preprint,preview,12pt]{elsarticle}

\usepackage{graphics}
\usepackage{graphicx}

\usepackage{amssymb}
\usepackage{amsmath}

\usepackage{latexsym}
\usepackage{makeidx}
\usepackage{multirow}
\usepackage{multicol}
\usepackage{float}
\usepackage{subfigure}
\usepackage{color, colortbl}
\usepackage{xcolor}

\usepackage{rotating}
\usepackage{caption}

\biboptions{square,numbers, comma, sort&compress}

\begin{document}

\begin{frontmatter}



\title{Evaluation of the Impact of Low-Emission Zone: \textbf{Madrid Central} as a Case Study.}


\author[mcm]{Miguel C\'ardenas-Montes}
\address[mcm]{CIEMAT, Department of Basic Research. \\Avda. Complutense 40. 28040. Madrid, Spain.}
\ead{miguel.cardenas@ciemat.es}

\begin{abstract}
Air-quality in urban areas and its relation with public health is one of the most critical concerns for governments. 
For this reason, they implement actions aimed at reducing the concentration of the most critical atmospheric pollutants, among others the implementation of Low-Emission Zones.
\textbf{Madrid Central} is a major initiative of Madrid city council for reducing motor traffic and the associated air pollution at the city center. It is created by enlarging a previous and smaller area of restricted motor traffic. This initiative starts at the end of 2018, but the first fully-operational period corresponds to 2019Q2 (second quarter of 2019). 
The activation of Low-Emission Zones has social and economic impacts, besides environmental ones.
For this reason, their impact must be assessed to ensure the achievement of the commitments in the reduction of atmospheric pollutants levels.  
In this work, two metrics for evaluating the impact of \textbf{Madrid Central} in the reduction of nitrogen dioxide concentration are proposed.  
Mean daily concentrations of $NO_2$ from 2010 to 2019 corresponding to the \textit{Plaza del Carmen} monitoring station are employed for the evaluation of the impact of \textbf{Madrid Central}. 
Two other monitoring stations around \textbf{Madrid Central} are analysed for ascertaining the rise of detrimental effects due to an increment of the motor traffic around the traffic restricted area.  
In consequence, two methodologies pollutant-independent for evaluating the impact and the evolution of LEZs, as well as a statement about the impact of the first year of \textbf{Madrid Central} are presented.  
\end{abstract}

\begin{keyword}
LEZ evaluation \sep Metric \sep Madrid Central \sep Air Quality

\end{keyword}
\end{frontmatter}


\section{Introduction}


Air pollution is one of the most critical health issues in urban areas, being an important concern for citizens and governments. The scientific literature shows its relation with the population health \cite{AlberdiOdriozola1998,Diaz1999,Nel804,Linares2006,doi:10.1080/10473289.2006.10464485,doi:10.3109/08958378.2011.593587,BoogaardErp,India,ERivas,JMSantamaria}. 

Today, a large population of the World lives in urban agglomerations. In the next few years, the tendency of the distribution of the population will make these areas grow more while reducing the rural population. This trend will aggravate the problems caused by poor air quality if mitigating measures are not taken.

The concerns about air quality have led authorities to take various kinds of actions: from the establishment of networks of monitoring the level of multiple pollutants: $CO$, $NO_2$, $O_3$, $PM2.5$, $PM10$, toluene, etc; to energy efficiency policies in buildings and the implementation of traffic restrictions in areas with low-quality air as well as over the most polluting vehicles.

Motor traffic is a major source for $NO_2$ \cite{BORGE20181561}. German Environment Agency estimates that road transportation is responsible for 60\% of emissions of $NO_2$ in cities \cite{EBannon}.   
Thus, the establishment of Low-Emission Zones (LEZ) with restrictions on combustion motor vehicle traffic is frequently adopted in all major cities. They are usually implemented in the city center, for being in these areas where the air quality levels are lower. Following \cite{EBannon}, about 250 European cities have implemented LEZs.

The implementations of the LEZs are not without controversy. Contrary opinions are affected by the potential negative economic impact of the traffic restrictions within the LEZs while questioning their efficiency in the reduction of the levels of pollutants. Some of these considerations not only affect the impact on the inner area of the LEZ, but negative effects on the LEZ border are also mentioned. This effect would be due to an increment of the traffic around the LEZ, and therefore of the levels of pollutants. 
For this reason, the evaluation of the impact of the LEZ is mandatory to ensure the achievement of the goals in the reduction of pollutants concentration inside their area and the avoidance of any increment in their frontier. This requires the creation of the appropriate metrics.

After a year of operation of \textbf{Madrid Central}, and with the previous historical series, there is sufficient data for the evaluation of its impact on the levels of pollutants, and the effect on the border of the LEZ.

In this work two methodologies for evaluating the impact of the LEZ are proposed. The first one is based on the use of the statistical tests: the Binomial Sign Test for a Single Sample and the Chi-square Test for Homogeneity. They allow ascertaining whether the concentration is significantly low after the activation of \textbf{Madrid Central} or not. Besides, these tests are also applied to the wind and the rain ---as major pollution removers--- in the same period. 
This permits discarding any reduction of concentration in the concentration of the pollutants coming from significantly favourable meteorological scenarios. 
This methodology provides an indicator if the differences between the observations of two quarters ---before and after the activation of LEZ--- are significant or not.

The second methodology is based on the use of Gaussian Mixture Models (GMM). It uses as input the means and standard deviations from the fitting of the quarterly observations from 2010 to 2018 to a Gaussian probability distribution. Once the Gaussian distribution has been adjusted to these points, the likelihood of the corresponding point coming from the quarterly observations of 2019 is calculated. Later, the divergence of Jensen-Shannon is calculated between the point representing the observations of 2019 and the centroid of the Gaussian distribution. 
This methodology serves to measure the distance between the observations after the LEZ activation and a model of the observations of years before the LEZ, particularly from 2010 to 2018.

The proposed metrics are applied to the evolution of the concentration of Nitrogen Dioxide, $NO_2$, in the \textit{Plaza del Carmen} monitoring station. This monitoring station is the only one inside of \textbf{Madrid Central}. $NO_2$ is a critical pollutant during high-pressure winter anticyclone periods. In those periods, its concentration excesses the annual mean concentration recommended by WHO and the air quality standards of the European Union legislation  (see Section \ref{section:dataset}). Beside, $NO_2$ is one of the precursors of ground level Ozone in presence of sunlight \cite{ozone-basic}.

%

Concerning the Sustainable Development Goals (SDGs), this work may acting as enabler for the goals Goal 3 "Ensure healthy lives and promote well-being for all at all ages" and particularly for the Target 3.9 "By 2030, substantially reduce the number of deaths and illnesses from hazardous chemicals and air, water and soil pollution and contamination" \cite{SDG}. An adequate evaluation of the LEZ allows implementing the corrective action for those with an impact lower than expected in the reduction of air-pollutant concentration.

The main contributions of this work are:
\begin{itemize}
\item The proposed methodologies allow evaluating for periods shorter than one year the significance of the reduction of $[NO_2]$ in not favorable meteorological scenarios.
\item The methodology based on the Binomial Sign Test for a Single Sample and the Chi-square Test for Homogeneity is able to discriminate the influence of the meteorology. 
\item The Jensen-Shannon divergence allows studying the long-term evolution of air pollutants in the LEZ.
\end{itemize}

The rest of the document is organized as follows: 
in Section \ref{section:PreviousEfforts}, the previous efforts on the evaluation of the impact of LEZ are shown.  
Section \ref{section:MM} presents the main background of the work, including a description of the dataset, the Binomial Sign Test for a Single Sample, the Chi-square Test for Homogeneity, the Gaussian Mixture Models algorithm, and the Jensen-Shannon Divergence. 
The Results and the Analysis are shown in Section \ref{section:results}. Finally, the Conclusions are presented in Section \ref{section:conclusions}. 

\section{Previous Efforts on the Evaluation of LEZ\label{section:PreviousEfforts}}

Report \cite{EBannon} reviews the evidence available in the previous bibliography on the justification and effectiveness of LEZ in diverse geographical areas: Germany, Denmark, Netherlands, Italy, UK, etc.; and cities: Lisbon, Brussels, London, Milan, and Madrid. In this report, the activation of \textbf{Madrid Central} is claimed as "The highest reduction observed is a decrease of the NO2 concentration in Madrid by 32\%." in this report.

Also in \cite{EBannon} a body of evidence for the reduction of pollutants in the implemented LEZs, not all of them scientific publications, are cited. The case of \textbf{Madrid Central} is mentioned, corresponding the analysis to the year 2019 and in turn based on \cite{ecologistasenaccion_MC_y1}. 
In this report, the analysis is based on the comparison of the annual mean concentration of $NO_2$ in 2019 versus the annual mean of the years from 2010 to 2018, and versus the mean of the period 2010-2018. In the first case, the mean of 2019, 36 $\frac{\mu g}{m^3}$ is lower than the annual mean concentration of the previous years 52 $\frac{\mu g}{m^3}$ for 2010 and 2011, 44 $\frac{\mu g}{m^3}$ for 2012, 41 $\frac{\mu g}{m^3}$ for 2013, 40 $\frac{\mu g}{m^3}$ for 2014, 50 $\frac{\mu g}{m^3}$ for 2015, 46 $\frac{\mu g}{m^3}$ for 2016, 49 $\frac{\mu g}{m^3}$ for 2017, and 45 $\frac{\mu g}{m^3}$ for 2018. The mean concentration of $NO_2$ for 2019 is also lower than the mean of the period 2010-2018, 46 $\frac{\mu g}{m^3}$, a 22\% lower.

The number of days of the period 2016-2019 for which the protocol of high concentration of $NO_2$ has been activated is also used as an indicator of the impact of \textbf{Madrid Central} in this report. For this indicator the number of days is 14 days for 2019, versus 11 days for 2018, 33 days for 2017, and 10 days for 2016. 

Influence of the meteorological conditions is analysed ---through the number of rainy days--- for evaluating their influence in the reduction of $NO_2$ concentration. In 2019, 76 rainy days, with an annual mean rainfall of 391.5 $\frac{l}{m^3}$, are accounted versus a mean of 86 rainy days the period 2010-2018, with an annual mean rainfall of 407.7 $\frac{l}{m^3}$.

In \cite{doi:10.1080/10962247.2013.868380}, the difficulty to factorise the reductions observed in percentage in some pollutants after the LEZ implantation from the influence of a favourable or unfavourable meteorology is analysed. Also the evaluation of the LEZ for periods shorter than a year with the adequate removal of the seasonal biases is highlighted. The authors advise an assessment of the LEZ contribution in terms of the air quality of a year.

In \cite{CHolman}, authors underline the difficulty associated to an exhaustive evaluation of the impact of the LEZ implantation: "There have been relatively few studies which have attempted to evaluate the impact of a LEZ using measured concentrations, possibly because of the difficulty in identifying small changes in concentrations following policy interventions.". This work makes a review of the LEZ implanted in five EU countries (Denmark, Germany, Netherlands, Italy and UK), including aspects of the date of implantation, the promotion of cleaner transport or the banning of the very pollutant vehicles, and the modeling and monitoring studies. Concerning the latter, except for \cite{Morfeld}, in most of the cases the analyses are based on the calculation of the percentage of reduction of the most critical pollutant. 

In \cite{AnnaFont}, the measured exceedances of the European Limit Value (ELV) for $NO_2$ (40 $\frac{\mu g}{m^3}$ annual mean) and for PM10 (40 $\frac{\mu g}{m^3}$ annual limit) before and after the activation of the LEZ in Paris and London are used as criteria of the impact of the LEZ activation. Multiple monitoring stations inside the LEZ are included in the study.

Finally, in \cite{Morfeld} a linear regressor on the paired differences between the same day before and after the LEZ implementation is created. This regressor incorporates some terms with the effect of meteorological conditions. The statistical significance of the coefficients of each term of the linear regressor is analysed.  

In \cite{10.1007/978-3-030-41913-4_10} the impact of \textbf{Madrid Central} is evaluated through the analysis of $NO_2$ time series with polynomial regression and Recurrent Neural Network (RNN). Hourly $NO_2$ concentrations at \textit{Plaza del Carmen} station are used. Polynomial regression models and a RNN are trained with time series concentrations before the activation of \textbf{Madrid Central}. Later two periods are evaluated with these models, before and after activation. In each period, the percentage of predictions over the real values is used as an indicator of the impact of the LEZ. On the one hand, polynomial regressors are unable to capture the hourly, weekly and annual modulations. And, on the other hand, RNN accuracy can not clearly be linked to the impact of the LEZ activation. 
Finally, the impact of a potential LEZ for Pamplona (Spain) is presented in \cite{JMSantamaria}. Also the negative impact of the increment of traffic around the LEZ is discussed. 

A different question is which is the most appropriate pollutant for evaluating the impact of the LEZ. Authors of the cited literature indicate either PM2.5, PM10, $NO_x$ and $NO_2$. Motor traffic has a larger contribution to $NO_x$ and $NO_2$ than to PM2.5 and PM10.
In any case, PM10 and PM2.5 are not pollutants included in the \textit{Plaza del Carmen} monitoring station, thus the choice for our analysis is $NO_2$. At this point, an exception is \cite{10.1007/978-3-030-38889-8_2} which uses the noise level as well as the $NO_2$ concentration for evaluating the impact of a LEZ.

\section{Methods and Materials\label{section:MM}}

\subsection{Madrid Central Low-Emission Zone}

Officially \textbf{Madrid Central} is inaugurated on 30 November 2018 \cite{elmundo_20181129,elpais_20181130}. From this date until 16 March 2019 drivers circulating without authorisation receive a notification but no sanctions. At a later date the sanction period is activated. It has an area of 472 hectares. The area enclosed by this LEZ has very little industrial activity, and the majority of pollutant sources corresponds to heating systems and motor traffic. \textit{Plaza del Carmen} is the only air-quality monitoring station inside of \textbf{Madrid Central}. Only residents and guests, electric and hybrid cars can access to the inner of the LEZ. Diesel and gasoline vehicles are granted a moratorium until 2020.

In July 2019, the new local government established a moratorium on the sanctions of \textbf{Madrid Central}, which in fact imply the suspension of the LEZ. One of the arguments is based on the lack of an effective reduction of the pollutant concentrations. This dismantlement was initially suspended and later annulled by two sentences in June 2020 \cite{eldiario_20200617}. This motivates to in-detail assess the first year of \textbf{Madrid Central}.

\subsection{The Dataset\label{section:dataset}}
Data from Air Quality Monitoring Network of Madrid are publicly available \cite{opendatamadrid}. It offers hourly and daily data from more than 24 monitoring stations, including three categories: suburban (stations in parks in urban areas), traffic (term for stations affected by traffic and close to a principal street or road), and background (urban background station affected by both traffic and background pollution). 

\begin{figure*}
{\renewcommand{\arraystretch}{1.0}
\rotatebox{90}{
\begin{minipage}[c][][c]{\textheight}
\centering
  \subfigure[Q1]{
    \includegraphics[width=0.482\textwidth]{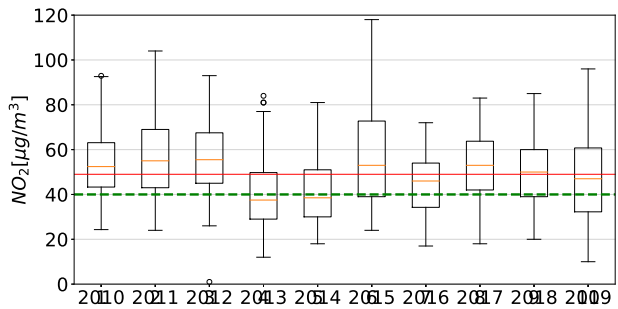}
  \label{figure:boxplot:2019Q1:NO2:PlC}}
  \subfigure[Q2]{
    \includegraphics[width=0.482\textwidth]{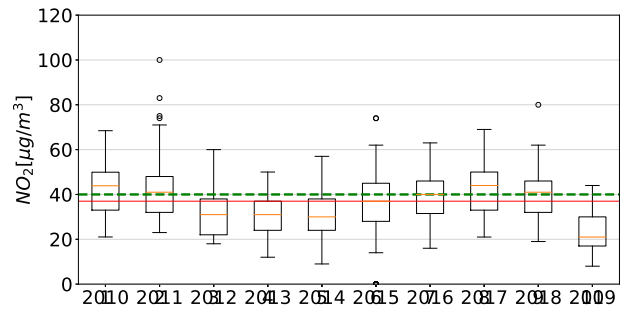}
  \label{figure:boxplot:2019Q2:NO2:PlC}}
  \subfigure[Q3]{
    \includegraphics[width=0.482\textwidth]{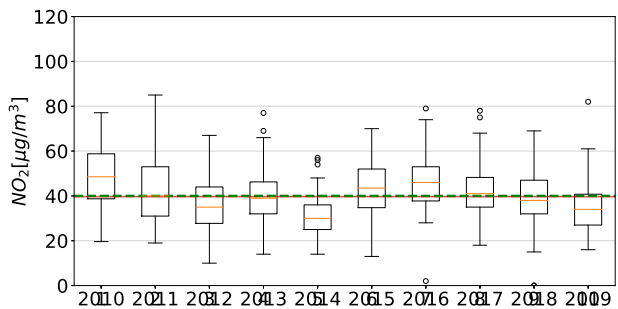}
  \label{figure:boxplot:2019Q3:NO2:PlC}}
  \subfigure[Q4]{
    \includegraphics[width=0.482\textwidth]{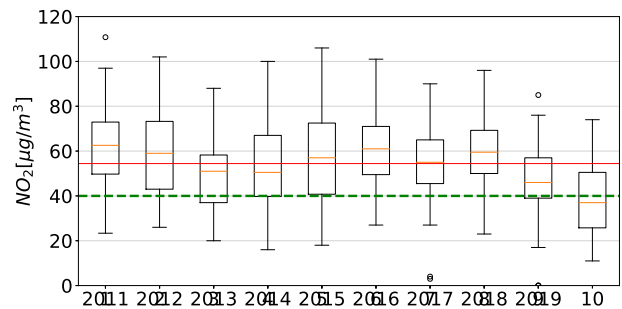}
  \label{figure:boxplot:2019Q4:NO2:PlC}}
  \caption{Boxplot with the mean daily concentration of $NO_2$ for the periods Q1 (Fig. \ref{figure:boxplot:2019Q1:NO2:PlC}), Q2 (Fig. \ref{figure:boxplot:2019Q2:NO2:PlC}), Q3 (Fig. \ref{figure:boxplot:2019Q3:NO2:PlC}), and Q4 (Fig. \ref{figure:boxplot:2019Q4:NO2:PlC}) at \textit{Plaza del Carmen} monitoring station. Red horizontal line shows the median of the values for the period 2010-2018. The green dashed line corresponds to the WHO guideline for the $NO_2$ annual mean concentration 40 $\frac{\mu g}{m^3}$ \cite{WHO}.} 
\label{figure:boxplot:2019:NO2:PlC}
\end{minipage}
}
}
\end{figure*}

In Fig. \ref{figure:boxplot:2019:NO2:PlC}, the boxplots with the mean daily concentration of $NO_2$ at \textit{Plaza del Carmen} monitoring station for the quarters from 2010 to 2019 are shown.  The red horizontal line shows the median of the values for the period 2010-2018. The green dashed line corresponds to the WHO guideline for the $NO_2$ annual mean concentration 40 $\frac{\mu g}{m^3}$ \cite{WHO}.


In November 2018 \textbf{Madrid Central} was started, however, until 2019Q2, only notifications were sent to drivers entering the restricted area without the adequate permission. As it can be observed, median of 2019Q1 measurements (46.5 $\frac{\mu g}{m^3}$) is slightly below of the median of the values for the period 2010-2018 (49.0 $\frac{\mu g}{m^3}$), but above of the WHO annual recommendation (Fig. \ref{figure:boxplot:2019Q1:NO2:PlC}). 

For the period 2019Q2, the first fully operational quarter, a reduction of mean daily concentration of $NO_2$ is appreciated (Fig. \ref{figure:boxplot:2019Q2:NO2:PlC}). Comparing with the median of this quarter for the period 2010-2018, 37.0 $\frac{\mu g}{m^3}$; the median of 2019Q2 is much lower, 21.0 $\frac{\mu g}{m^3}$. This indicates a clear reduction of the concentration of $NO_2$ in this station.  

The period 2019Q3 (Fig. \ref{figure:boxplot:2019Q3:NO2:PlC}) corresponds to the summer period, for which the sources of $NO_2$ are the lowest: due to high temperature, heating systems are usually off; and due to the summer holidays, motor traffic is critically reduced at Madrid. Also, solar irradiation removes part of the $NO_2$ by dissociating it as a precursor of ozone in the troposphere. 
Visually, the mean daily observation for 2019Q3 are not much lower (34 $\frac{\mu g}{m^3}$) than the same period in the previous years (40 $\frac{\mu g}{m^3}$).

Finally, for the period 2019Q4 (Fig. \ref{figure:boxplot:2019Q4:NO2:PlC}), a strong reduction of the $NO_2$ concentration is appreciated (37 $\frac{\mu g}{m^3}$) compared with the previous years (54 $\frac{\mu g}{m^3}$). This period includes winter months for which some of the most critical days appear. These days coincide with winter high pressure days, without wind and rain, and formation of low level thermal inversion close to the surface avoiding the turbulence and pollutant dispersion.  

Beyond the visual inspection of the daily mean values, the objective of the current work is to apply the appropriate metrics for ascertaining if the observed reductions are significant or not from the statistical point-of-view; and to evaluate the distance between the observations coming from active periods of LEZ and previous periods.

\subsection{The Binomial Sign Test for a Single Sample\label{section:BinomialSignTest}}
The Binomial Sign Test for a Single Sample is based on the binomial distribution (Eq. \ref{eq:BinomialDistribution}) \cite{Sheskin2004}. It assumes that any observation can be classified in one of the two mutually exclusive categories with probabilities $\pi_1$ and $\pi_2$. The evaluated hypothesis is if the proportion of the two categories are equal to a specific value, usually if they are equal. 




\begin{equation}
P(x)= \frac{n!}{x!(n-x)!} \, \pi^x_1 \, \pi^{n-x}_2
\label{eq:BinomialDistribution}
\end{equation}
where $\pi_1 + \pi_2=1$, and therefore, $\pi_2=1-\pi_1$; $n$ is the sample size or the number of observations, and $x$ is the number of positive signs ($n-x$ is the number of negative signs).

The two-tailed Null Hypothesis assumes that the true proportion of observations in any of the two categories are equal to $0.5$, $H_0: \, \pi_1=0.5$. Conversely, the two-tailed Alternative Hypothesis assumes that this is not true: \linebreak ${H_0: \, \pi_1 \ne 0.5}$.

In this work, the two categories correspond to the sign of the differences when subtracting the daily mean concentration of $NO_2$ (Eq. \ref{eq:pairwisecomp}). Each sample includes the total daily mean values for Q1, 90 days; 91 days for Q2\footnote{In order to maintain an equal number of samples in the second quarter, the measurements corresponding to the 29th of February in leap years are removed.}, and 92 days for Q3 and Q4. Samples are generated for each quarter of year: Q1, Q2, Q3 and Q4; and for all the pairwise comparisons between a quarter of 2019 and a quarter of the years in the period 2010-2018. Thus, if the daily mean values of Q1 for 2019 are significantly lower than the values of the other year of the comparison, then an excess of positive differences are obtained, $\pi_1>>0.5$. Conversely, if the daily mean values of 2019 are significant larger than the values of the other year of the comparison, then the excess corresponds to negative differences, $\pi_1<<0.5$. Finally, if the two values sets involved in the comparison behave similarly, then the probability of positive signs and negative signs will be similar and close to $0.5$. In this case the confidence interval for $\pi_1$ includes $\pi_1=0.5$.

\begin{eqnarray}
& [NO_2]^{2010-01-01}-[NO_2]^{2019-01-01} >0 := + \nonumber\\ 
& [NO_2]^{2010-01-02}-[NO_2]^{2019-01-02} <0 := - \nonumber\\ 
& [NO_2]^{2010-01-03}-[NO_2]^{2019-01-03} <0 := - \nonumber\\ 
& [NO_2]^{2010-01-04}-[NO_2]^{2019-01-04} >0 := + \nonumber\\ 
& \vdots \nonumber\\ 
& [NO_2]^{2010-31-03}-[NO_2]^{2019-31-03} <0 := - 
\label{eq:pairwisecomp}
\end{eqnarray}


The computation of the confidence interval for $\pi_1$ is executed with the Eq. \ref{eq:BinomialDistributio:CI}.

\begin{equation}
p_1 - z_{\alpha/2} \sqrt{\frac{p_1 p_2}{n}} \le \pi_1 \le p_1 + z_{\alpha/2} \sqrt{\frac{p_1 p_2}{n}} 
\label{eq:BinomialDistributio:CI}
\end{equation}
where $p_1$ and $p_2$ are the proportion of positive and negative differences. For a confidence interval of 95\%, it must be computed through Eq.  \ref{eq:BinomialDistributio:CI95}.

\begin{equation}
p_1 - 1.96 \cdot \sqrt{\frac{p_1 p_2}{n}} \le \pi_1 \le p_1 + 1.96 \cdot \sqrt{\frac{p_1 p_2}{n}} 
\label{eq:BinomialDistributio:CI95}
\end{equation}

When the confidence interval does include the value $0.5$, it will mean that the differences are not significant for a confidence level of 95\% (p-value under 0.05). In this case, the Null Hypothesis can not be rejected, and therefore, no impact can be attributed to the LEZ activation. 

The Binomial Sign Test for a Single Sample has been implemented in R with package MASS \cite{VenablesRipley}. It allows compute the confidence interval for $\pi_1$. 

This methodology is not only applied to the $NO_2$ concentration measured at \textit{Plaza del Carmen} monitoring station, but also to the daily mean wind velocity, measured at Adolfo Su\'arez Madrid-Barajas Airport by AEMET (Spanish State Meteorological Agency). Wind and rain are efficient pollution removers and for this reason their activity is also evaluated. The application of statistical tests to the wind velocity and the number of rainy days allow discerning if similar meteorological conditions stand for the two periods under comparisons. 

For a meteorological scenario with not similar conditions between the two periods, the reduction of the pollution can be caused by the LEZ activation, or by the presence of more windy and rainy days in the period analysed, or a combination of both.  If a significant excess of windy and rainy days is observed during the comparison of a period of 2019, then any significant reduction of $NO_2$ concentration for this period can not be assigned to the LEZ activation, and therefore, it should be discarded as exclusively caused by the LEZ activation. 


In the previous efforts, this methodology was firstly tested for analysing a potentially harmful effect in the border of \textbf{Madrid Central}. It was used for evaluating the increment of PM2.5 in \textit{Escuelas Aguirre} monitoring station \cite{crdenasmontes2020report}.

Also the effect of the rain as pollution remover is analysed. In this case the Chi-square Test for Homogeneity is employed.  In the Section \ref{section:Chi2Test}, the reasons for the use of this test instead of the Binomial Sign Test for a Single Sample. 
Similarly the wind analysis, if a significant excess of rainy days is observed in the quarters of the year 2019, then any potential reduction of the $NO_2$ concentration can not be identified to the LEZ activation.

\subsection{The Chi-square Test for Homogeneity\label{section:Chi2Test}}
The Chi-square Test for Homogeneity is based on the chi-square distribution. 
It is employed when independent samples are categorised on a single dimension which consists of categories, and it evaluates whether or not the samples are homogeneous with respect to the proportion of observations in each of the categories \cite{Sheskin2004}. Samples in the categories are represented as a contingency table.

Since there could be many days without rain in a quarter (see Table \ref{table:railfall}), and as a consequence the array of precipitation can contain many null elements, the Binomial Sign Test for a Single Sample seems not suitable. For this reason, the Chi-square Test for Homogeneity is applied to the number of rainy days in the quarter.

\begin{table}
\caption{Number of rainy days in the quarters with precipitation larger than 0.1 $\frac{l}{m^3}$ at Adolfo Suárez Madrid-Barajas Airport station.}
\label{table:railfall}
\centering 
\begin{tabular}{crrrr} \hline
Year & Q1 & Q2 & Q3 & Q4 \\ \hline \hline
2010 & 45 & 27 &  6 & 28 \\
2011 & 33 & 22 &  6 & 23 \\
2012 &  6 & 23 & 10 & 26 \\
2013 & 37 & 22 &  8 & 19 \\
2014 & 32 & 18 &  7 & 17 \\
2015 & 17 & 11 &  7 & 17 \\ 
2016 & 29 & 30 &  4 & 28 \\
2017 & 20 & 16 &  7 & 10 \\
2018 & 31 & 37 &  6 & 37 \\
2019 &  7 & 13 & 14 & 35 \\ \hline
\end{tabular}
\end{table}

This test assumes that the categories are mutually exclusive, that the data which are evaluated represent a random sample consisting of independent observations, and that the expected frequency of each cell in the contingency table is 5 or greater. A less conservative criterion is stated by Cochran (1952), for which none of the expected frequencies should be less than 1, and that no more than 20\% of the expected frequencies should be less than 5 \cite{Cochran}. 

The Null Hypothesis of the Chi-square Test for Homogeneity assumes that for each of the cells in the contingency table the observed frequency is equal to the expected frequency of the cell. Oppositely the Alternative Hypothesis assumes that, for at least one cell in the contingency table the observed frequency of the cell is not equal to the expected frequency.

Alternative way of stating the null and alternative hypotheses for the Chi-square Test for Homogeneity is that the underlying populations the samples represent, all of the proportions in the same column of the contingency table are equal. And the alternative hypothesis states that all of the proportions in the same column of the contingency table are not equal for at least one of the columns.

In this work the test is used for evaluating the homogeneity in the proportion of rainy and not rainy days in a quarter between two years under comparison.  For this purpose, a $2 \times 2$ contingency table involving the rainy days and not rainy days of the quarter of years is created and evaluated with the test. This process is repeated for all the pairs between a quarter of 2019 and the same quarter of the years in the period from 2010 to 2018. In Table \ref{table:railfall}, the number of rainy days with rainfall larger than 0.1 $\frac{l}{m^3}$ are shown. An example of a contingency table is shown in Table \ref{table:contingencytable}. 

\begin{table}
\caption{Example of contingency table for the comparison between 2010Q1 and 2019Q1. Q1 period has 90 days.}
\label{table:contingencytable}
\centering 
\begin{tabular}{crr} \hline
Period & Rainy Days & Not Rainy days  \\ \hline \hline
2010Q1 & 45 & 90-45=45 \\
2019Q1 &  7 & 90-7=63 \\ \hline
\end{tabular}
\end{table}

Although it is possible to apply a directional alternative hypothesis for the Chi-square Test for Homogeneity to the contingency tables, in this work, the previously-described non-directional version is employed. Therefore, only the pairwise comparisons with more rainy days in the 2019 quarter than in the under comparison quarter with statistically significance are retained. 

As efficient pollution remover, the rain, through the number of rainy days, must be evaluated in order to discard beneficial conditions for certain quarters of years, and finally ascertain if the reduction in the $[NO_2]$ quarter with the LEZ active can be definitively associated to this activation or an abnormal rainy quarter.  
As for the previous test, a confidence level of 95\% (p-value under 0.05) is used for rejecting the Null Hypothesis.

\subsection{Gaussian Mixture Models}
Gaussian Mixture Models (GMM) are probabilistic models constructed with the mixture of a set of Gaussian probability distributions (Eq. \ref{eq:GMM:terms}) \cite{DBLP:books/mk/HanK2000,Bishop:1995:NNP:525960}. The mixture is a weighted sum of terms. Each term of the sum is composed of a weight $w_i$, and a Gaussian function $N(\vec{x}|\mu_i, \Sigma_i)$.

\begin{equation}
\label{eq:GMM:terms}
p(\vec{x}|w_i,\vec{\mu_i},\Sigma_i)=\sum_{i=1}^M w_i\cdot N(\vec{x}|\vec{\mu_i}, \Sigma_i)
\end{equation}

The weights, $w_i$, correspond to the probability of the point $i$ to belong to the distribution $N(\vec{x}|\mu_i, \Sigma_i)$. The accumulated probability for any point to be in the set of distributions should be the unit, $\sum_{i=1}^M w_i=1$, where $w_i\ge 0$. Through $w_i$  meaning, each object has a certain probability to be a member of a given cluster.

Each component of the weighted sum is a Gaussian function with dimensionality of the problem's dimensionality (Eq. \ref{eq:GMM:gaussian}).
\begin{equation}
\label{eq:GMM:gaussian}
N(\vec{x}|\vec{\mu_i}, \Sigma_i) \,=\, \frac{1}{{(2\pi)}^{D/2} |\Sigma_i|^{1/2}} \cdot e^{-\frac{(\vec{x}-\vec{\mu_i})^T\Sigma_i^{-1}(\vec{x}-\vec{\mu_i})}{2}}
\end{equation}
where $D$ is the dimensionality of the problem, $\vec{\mu_i}$ is the vector of the mean of the distribution, and $\Sigma_i$ is the covariance matrix. The flexibility of GMM holds on the variety of covariances types: spherical, diagonal, full or tied. 

GMM has been frequently used for unsupervised learning and outlier detection. Their strengths include its high speed, and the low tend to be biased. Oppositely, among its weaknesses it can cited the need to declare the number of clusters, which requires a priori knowledge of the data.

Usually GMM appears associated with Expectation-Maximization (EM) algorithm for fitting the values of their parameters, $\left\{w_i,\vec{\mu_i},\Sigma_i\right\}$. EM is an iterative algorithm which after the random initialization follows two steps: Expectation and Maximization. During the Expectation step, each object $\vec{x}$ is assigned to a cluster $C_k$ with a certain probability which depends on the object and the cluster's parameters, $\left\{\vec{\mu},\Sigma\right\}$, (Eq. \ref{eq:EM:EStep}). Next, in the Maximization step, the weight of the objects are recalculated (Eq. \ref{eq:EM:MStep}).

\begin{equation}
\label{eq:EM:EStep}
P(\vec{x_i} \in C_k) = p(C_k|\vec{x_i}) = \frac{p(C_k)p(\vec{x_i}|C_k)}{p(\vec{x_i})}
\end{equation}

\begin{equation}
\label{eq:EM:MStep}
w_i=\frac{1}{n} \sum_{i=1}^n \frac{\vec{x_i}P(\vec{x_i} \in C_k)}{\sum_j P(\vec{x_i} \in C_j)}
\end{equation}

EM algorithm has a fast convergence, however the convergence to the global minimum is not guaranteed, it can fall in local minima. In order to avoid this flaw, usually the implementations make more than one start for selecting the most suitable minimum, and therefore, for approaching the global minimum.

By ending the process, a model for describing the data set is produced. This model provides a probabilistic value for the association of each object to each cluster. This information can be used for stating the probability of an object belonging to a certain cluster, or for labeling as outlier the objects with low-probability of belonging to all the clusters as outliers \cite{books/sp/Aggarwal2013}.

\subsection{Jensen-Shannon Divergence\label{section:JSd}}
The Jensen-Shannon divergence (JS) is the symmetric version of the Kullback-Leibler divergence (KL) (Eq. \ref{eq:KL:1}) \cite{JSd}. The Kullback-Leibler divergence aims at measuring how much a probability distribution differs from a second one \cite{kullback1951,kullback1959}. The KL divergence is not symmetric, $KL(p||q) \ne KL(q||p)$, and therefore, it is not a metric (see Section \ref{section:metrics} for a review of the requirements for a metric). This drawback is overcome by making symmetrical the role of both distributions: $JS = \frac{1}{2}(KL(p||q) + KL(q||p))$. 


\begin{equation}
KL(p||q) = \int p(x) \, log \left( \frac{p(x)}{q(x)} \right) \,dx
  \label{eq:KL:1}
\end{equation}

If the probability distributions $p(x)$ and $q(x)$ are Gaussian probability distributions, then the KL divergence takes the form of the Eq. \ref{eq:KL2gauss:6}.

\begin{align}
  \label{eq:KL2gauss:6}
KL(p||q) &=  \frac{1}{2} log\left( \frac{\sigma_2^2}{\sigma_1^2} \right) - \frac{1}{2} + \frac{1}{2\sigma_2^2} \left( \sigma_1^2 + (\mu_1-\mu_2)^2 \right) 
\end{align}
where $\mu_1$ and $\mu_2$ are the means of the Gaussian probability distributions, and $\sigma_1^2$ and $\sigma_2^2$ are the variances. 

\subsection{Distance Metrics\label{section:metrics}}
A key point in the analysis of the impact of the LEZ is to define the appropriate metrics for its evaluation. These metrics should allow both their evaluation and the study of evolution over the years. 

In \cite{DBLP:series/aikp/SimoviciD08} a dissimilarity metrics is defined by the following:

A dissimilarity on a set $S$ is a function $d: S^2 \rightarrow \mathbb{R}_{\ge 0}$ satisfying:
\begin{enumerate}
\item ($DIS_1$), $d(x,x)=0 \; \forall x \in S$; and
\item ($DIS_2$), $d(x,y)=d(y,x) \; \forall x,y \in S$.
\end{enumerate}

The properties of a such dissimilarity metrics are:
\begin{enumerate}
\item $d(x,y)=0$ implies $d(x,z)=d(y,z), \; \forall x,y,z\; in\; S$; \textbf{evenness} 
\item $d(x,y)=0$ implies $x=y, \; \forall x,y\; in\; S$; \textbf{definiteness}
\item $d(x,y) \le d(x,z) +d(z,y), \; \forall x,y,z$; \textbf{triangular inequality}
\end{enumerate}

The Jensen-Shannon divergence defined in Eq. \ref{eq:KL2gauss:6} fulfils these properties, and therefore it is employed for measuring the dissimilarity of point representing a period of 2019 and the centroid of the points representing the periods from 2010 to 2018, which in turn has been calculated with GMM with a single component.

\section{Results of \textbf{Madrid Central} and Discussion \label{section:results}}
In this section, firstly the impact of \textbf{Madrid Central} is evaluated by using Binomial Sign Test for a Single Sample. This allows ascertaining whether the variation of $NO_2$ concentration is significant or not. Next, the wind conditions ---as a major pollution remover--- during the same periods are also analysed with this statistical test. Also the number of rainy days with rainfall larger than 0.1 $\frac{l}{m^3}$ is analysed using the Chi-square Test for Homogeneity. With the results of both analyses, reduction of $NO_2$ concentration due to beneficial weather conditions can be pointed, and therefore, put aside.

Besides, the $NO_2$ is also analysed with GMM and the Jensen-Shannon divergence. This allows creating a metric for evaluating the distance between the air quality of a quarter of the year before and after the LEZ activation. 


\subsection{Binomial Sign Test for a Single Sample Applied to $[NO_2]$}

\begin{figure*}
\centering
  \subfigure[2019Q1]{
    \includegraphics[width=0.39\textwidth]{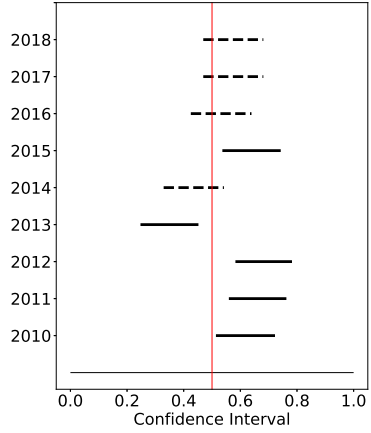}
  \label{figure:ci:binomial:PlC:2019Q1}}
  \subfigure[2019Q2]{
    \includegraphics[width=0.39\textwidth]{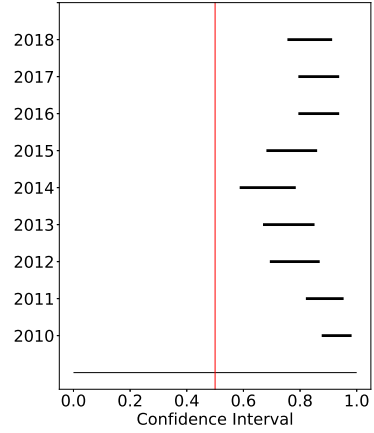}
  \label{figure:ci:binomial:PlC:2019Q2}}\\
  \subfigure[2019Q3]{
    \includegraphics[width=0.39\textwidth]{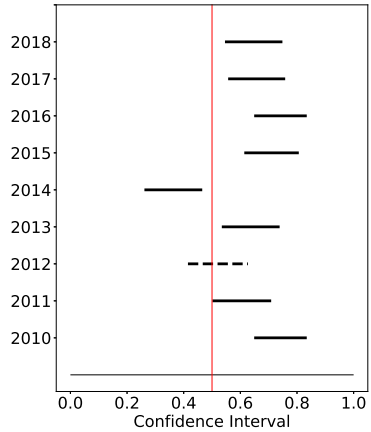}
  \label{figure:ci:binomial:PlC:2019Q3}}
  \subfigure[2019Q4]{
    \includegraphics[width=0.39\textwidth]{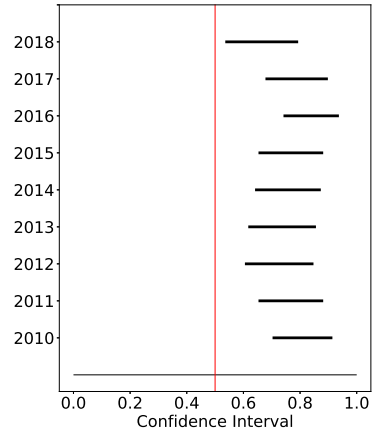}
  \label{figure:ci:binomial:PlC:2019Q4}}
\caption{Confidence intervals of $\pi_1$ for the pairwise comparisons with Binomial Sign Test for a Single Sample for $NO_2$ concentration at  \textit{Plaza del Carmen} monitoring station. The pairwise comparisons are undertaken by pairing the mean daily concentrations of 2019 versus the corresponding previous years. Solid lines indicate that the differences are significant, whereas dotted lines indicate that they are not significant.}
\label{figure:ci:binomial:PlC}
\end{figure*} 

In Fig. \ref{figure:ci:binomial:PlC:2019Q1}, the confidence intervals of $\pi_1$ for the pairwise comparisons of 2019Q1 are shown. As it can be appreciated in four cases the confidence interval includes the value 0.5, indicating that the differences in behaviour for the pairwise comparisons with these years are not significant: 2014, 2016, 2017, and 2018 (dotted lines). 

The differences with respect to the year 2013 are significant, being the values of 2013Q1 lower than the values of 2019Q1. And for this reason, the probability $\pi_1$ of sampling positive values is significantly lower than $0.5$. 

For the remaining years (2010, 2011, 2012, and 2015), the excess of positive differences for the subtraction of the concentration of $NO_2$, 20xx-2019, is significant for a confidence level of 95\% (p-value under 0.05). In these cases, the probability $\pi_1$ of sampling positive values is significantly lower than $0.5$, and therefore the observations of 2019Q1 are significantly lower than the observations of 2010Q1, 2011Q1, 2012Q1, and 2015Q1.

For the pairwise comparisons of the period 2019Q2 (Fig. \ref{figure:ci:binomial:PlC:2019Q2}), the confidence intervals of $\pi_1$ indicate that all the differences are significant for confidence level of 95\% (p-value under 0.05). This means that the differences are unlikely to have occurred by chance with a probability of 95\%. This result underlines the reduction in the concentration of $NO_2$ observed in the \textit{Plaza del Carmen} monitoring station during the period 2019Q2 (Fig. \ref{figure:boxplot:2019Q2:NO2:PlC}). If similar meteorological scenarios can be demonstrated, then it could be stated that \textbf{Madrid Central} has caused the reduction observed in 2019Q2. 

For the period 2019Q3, the pairwise comparisons indicate that all the differences, but 2012, are significant for a confidence level of 95\% (Fig. \ref{figure:ci:binomial:PlC:2019Q3}). The differences with respect to the year 2014 are significant, being the observations of 2014Q3 lower than the observations of 2019Q3. For the remaining years, the observations of 2019Q3 are significantly lower, for confidence level of 95\% (p-value under 0.05), than the observations of the pairwise comparisons. 
 
Finally for the period 2019Q4 (Fig. \ref{figure:ci:binomial:PlC:2019Q4}), all the confidence intervals of $\pi_1$ for the pairwise comparisons indicate that the observations of the period 2019Q4 are significantly lower, for a confidence level of 95\%, than the pairwise comparisons. 

In three of the four periods analysed: 2019Q2, 2019Q3, and 2019Q4, the confidence intervals of $\pi_1$ are in favour of significant reductions of the mean daily concentration of $NO_2$ after the activation of \textbf{Madrid Central}. Unfortunately the role of the weather in these reductions has not been taken into account.

\subsubsection{Wind-based Weather Analysis}
A major remover of air pollution in large urban areas is the wind. Therefore, it must be ascertained if the reduction observed in the concentration of $NO_2$ are due to more windy days in 2019 quarters in comparison with the quarters of the previous years. For this purpose, the Binomial Sign Test for a Single Sample can also be used, but analysing the wind velocity observations. 

\begin{figure*}
\centering
  \subfigure[2019Q1]{
    \includegraphics[width=0.39\textwidth]{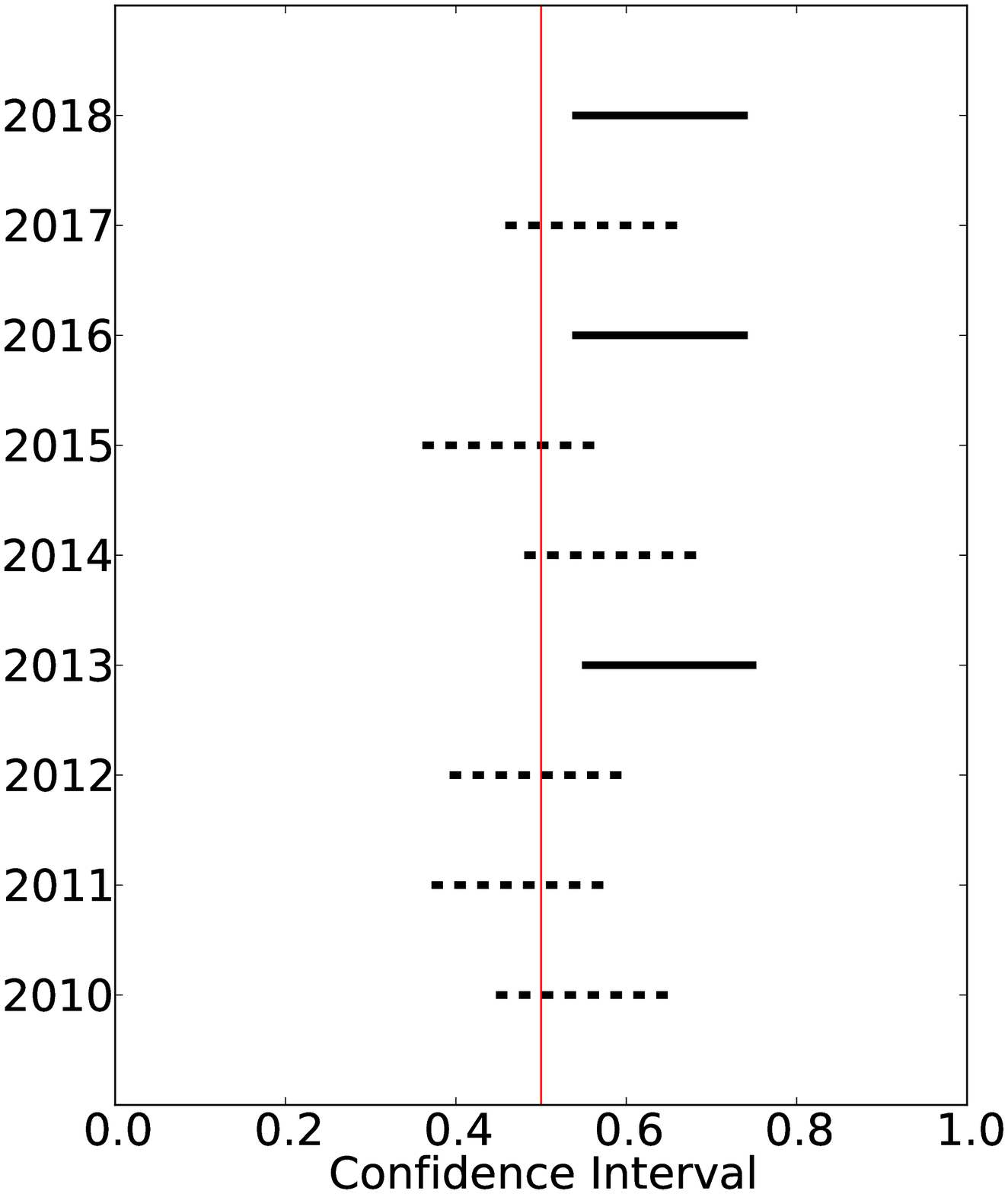}
  \label{figure:ci:binomial:wind:2019Q1}}
  \subfigure[2019Q2]{
    \includegraphics[width=0.39\textwidth]{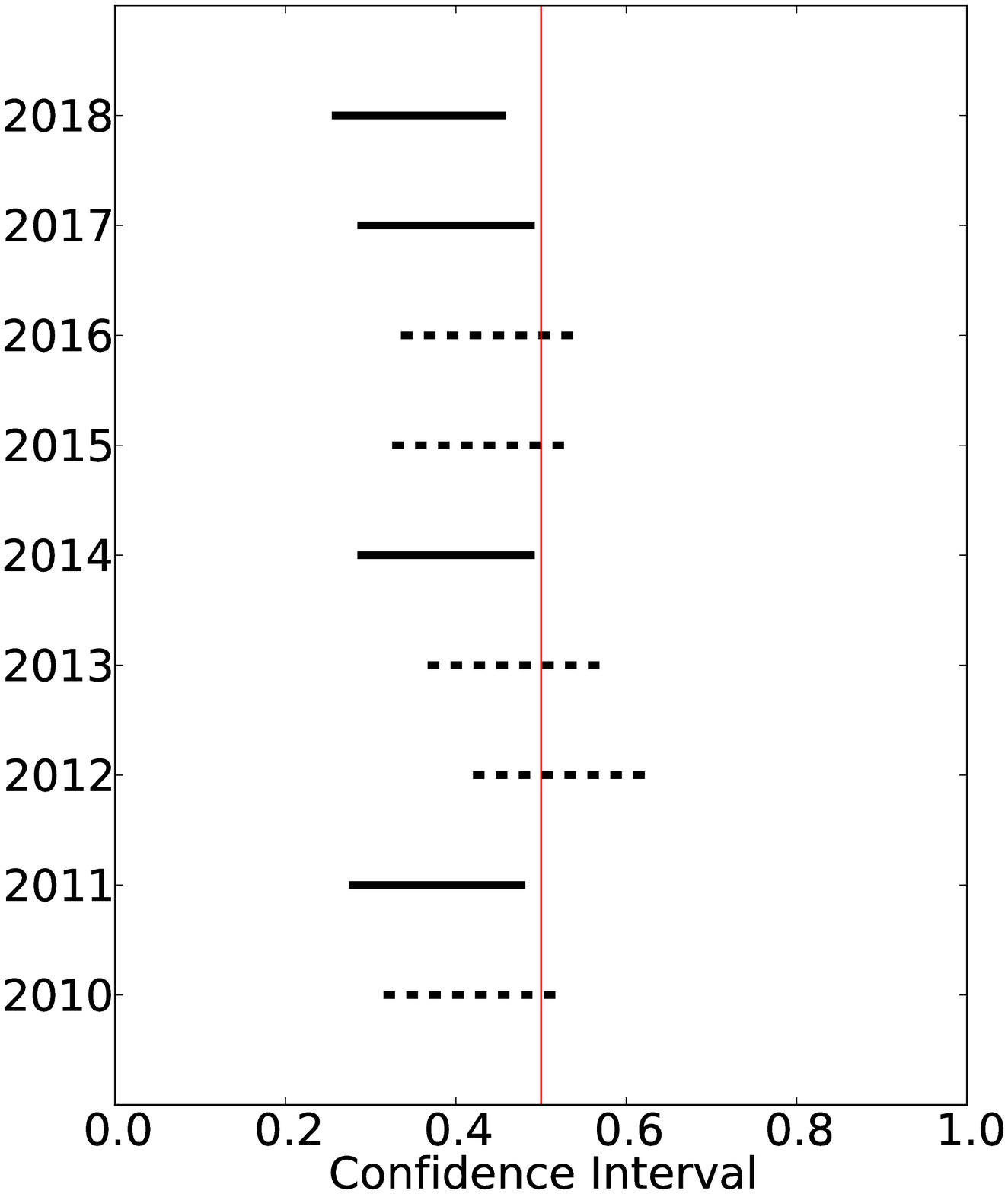}
  \label{figure:ci:binomial:wind:2019Q2}}
  \subfigure[2019Q3]{
    \includegraphics[width=0.39\textwidth]{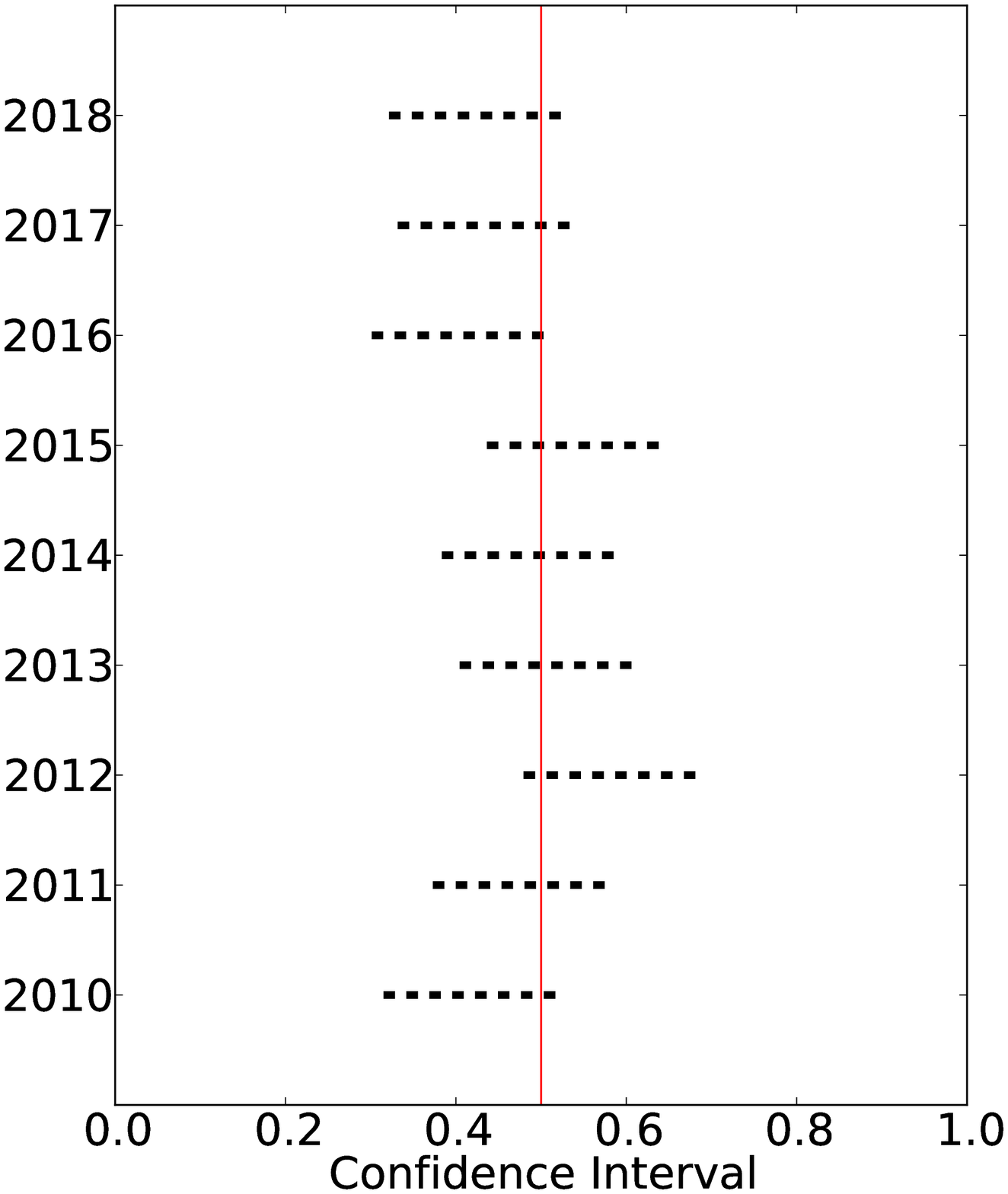}
  \label{figure:ci:binomial:wind:2019Q3}}
  \subfigure[2019Q4]{
    \includegraphics[width=0.39\textwidth]{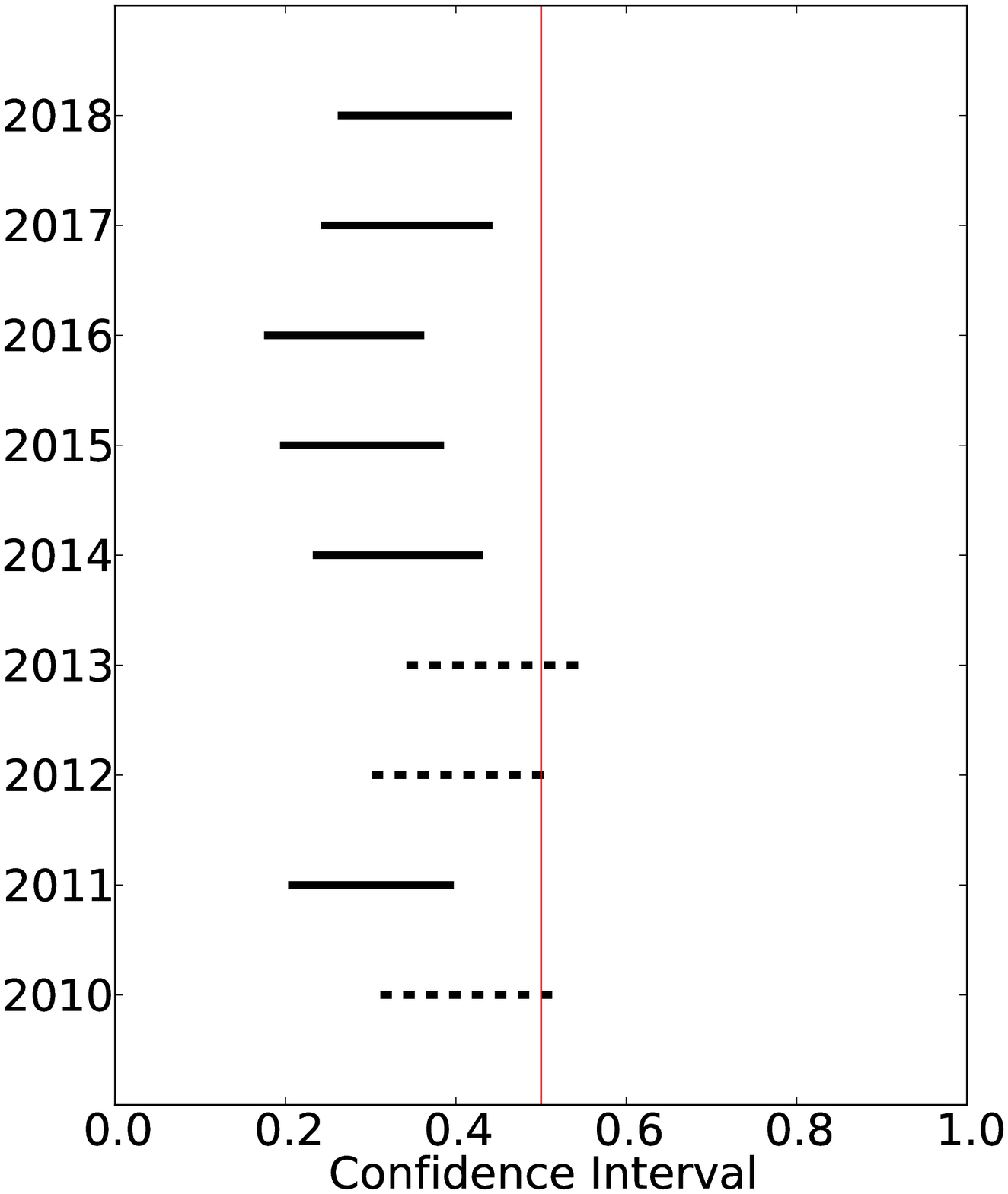}
  \label{figure:ci:binomial:wind:2019Q4}}
\caption{Confidence intervals of $\pi_1$ for the pairwise comparisons with Binomial Sign Test for a Single Sample for wind velocity at Adolfo Su\'arez Madrid-Barajas Airport. The pairwise comparisons are undertaken by pairing the mean daily observations of 2019 versus the corresponding previous years. Solid lines indicate that the differences are significant, whereas dotted lines indicate that they are not significant.}
\label{figure:ci:binomial:wind}
\end{figure*}

The pairwise comparisons are undertaken following the schema of Eq. \ref{eq:pairwisecomp} by using the wind velocity provided by AEMET at the Madrid-Barajas Airport station. More windy days 2019 quarters could lead to quarantine the reduction of concentrations in \textit{Plaza del Carmen} monitoring station due to a positive effect of \textbf{Madrid Central}. Oppositely, if the differences in the wind velocity observations do not show significant differences or the significant differences are in favour of less windy days in the 2019 quarters and reductions of $NO_2$ concentration are observed, then a positive effect due to the activation of the LEZ can be claimed\footnote{The definitive positive effect of LEZ activation only can be claimed after the analysis of full weather conditions, including the wind intensity and the number of rainy days.}. In Fig. \ref{figure:ci:binomial:wind}, the confidence intervals of $\pi_1$ for the pairwise comparisons of the wind velocity observations are shown.

Considering the first quarter analysis (Fig. \ref{figure:ci:binomial:wind:2019Q1}), no significant differences, for a confidence level of 95\%, are shown for the pairwise comparisons with the years: 2010, 2011, 2012, 2014, 2015, and 2017. Thus a similar meteorological scenario is stated for these years. 
Except for 2014 and 2017, for the rest of the years, significant reductions of the observations of $NO_2$ are also claimed (Fig. \ref{figure:ci:binomial:PlC:2019Q1}). 

For the second quarter of the year (Fig. \ref{figure:ci:binomial:wind:2019Q2}), no significant differences are shown for the pairwise comparisons with the years: 2010, 2012, 2013, 2015 and 2016; while positive reductions of $NO_2$ concentration are initially claimed for all the years. Now, only the positive reductions of $NO_2$ concentration of the years 2010, 2012, 2013, 2015 and 2016 ---similar meteorological scenarios--- can be assigned to the activation of the LEZ without a more favourable weather (Fig. \ref{figure:ci:binomial:PlC:2019Q2}). 

The analysis for the third quarter of the year indicates that the differences in the pairwise comparisons for windy days are not significant for any comparison (Fig. \ref{figure:ci:binomial:wind:2019Q3}). A similar meteorological scenario can be stated for all the years involved in the comparisons. Therefore, all the positive reductions of $NO_2$ concentration: 2010, 2011, 2013, and the years form 2015 to 2018 can be claimed linked to the LEZ activation without a more favourable wind regime (Fig. \ref{figure:ci:binomial:PlC:2019Q3}).

Except for the years: 2010, 2012 and 2013, in the fourth quarter of 2019 the days have been significantly more windy in most of the years comparisons (Fig. \ref{figure:ci:binomial:wind:2019Q4}). As a consequence, only for these years the reduction of $NO_2$ concentration can be associated to the LEZ activation with a similar meteorological scenario (Fig. \ref{figure:ci:binomial:PlC:2019Q4}).

The application of the Binomial Sign Test for a Single Sample has demonstrated its capacity to ascertain the significance of the reduction of $NO_2$ concentration. Furthermore, the application of this test is able to discard reduction cases when no wind conditions are pointed.  

The full analysis of the meteorological conditions requires the pairwise comparison of the number of rainy days using the Chi-square Test for Homogeneity.


\subsubsection{Rainfall-based Weather Analysis}
In Table \ref{table:chi2rain}, the p-values of the pairwise comparisons under the Chi-square Test for Homogeneity are shown. Only the cases with more rainy days in 2019 than in the comparison year (see Table \ref{table:railfall}) and with statistical significance ---confidence level of 95\% (p-value under 0.05)--- appear in bold face. For these cases, the potential reduction in the $NO_2$ concentration could not definitively be assigned to the activation of the LEZ.  

As it can be observed, the comparisons with 2016Q3, 2013Q4, 2014Q4, 2015Q4, and 2017Q4 indicate a significantly larger number of rainy days in the year 2019 for these quarters. The surviving $[NO_2]$ significant reductions pairwise comparisons are analysed in Section \ref{section:surviving}.

\begin{table}
\caption{P-values of the pairwise comparisons under the Chi-square Test for Homogeneity. The pairwise comparisons with more rainy days in 2019 than in the comparison year (see Table \ref{table:railfall}) and with statistical significance ---confidence level of 95\% (p-value under 0.05)--- are shown in boldface.}
\label{table:chi2rain}
\centering 
\begin{tabular}{crrrr} \hline
Year & Q1 & Q2 & Q3 & Q4 \\ \hline \hline
2018 & $2\cdot 10^{-5}$  & $1\cdot 10^{-4}$ & $0.10$ & $0.88$ \\ 
2017 & $0.011$ 		 & $0.69$  	& $0.16$ & $\mathbf{3\cdot 10^{-5}}$  \\
2016 & $6\cdot 10^{-5}$  & $5\cdot 10^{-3}$ & $\mathbf{0.02}$ & $0.35$ \\
2015 & $0.046$ 		 & $0.83$ 	& $0.16$ & $\mathbf{5\cdot 10^{-3}}$ \\
2014 & $8\cdot 10^{-6}$  & $0.43$ 	& $0.16$ & $\mathbf{5\cdot 10^{-3}}$ \\
2013 & $2\cdot 10^{-7}$  & $0.13$ 	& $0.25$ & $\mathbf{0.02}$ \\
2012 & $0.99$ 		 & $0.10$ 	& $0.51$ & $0.21$ \\
2011 & $4\cdot 10^{-6}$  & $0.13$ 	& $0.10$ & $0.08$ \\
2010 & $3\cdot 10^{-10}$ & $0.02$ 	& $0.10$ & $0.35$ \\ \hline
\end{tabular}
\end{table}

\subsubsection{Surviving $[NO_2]$ Reductions\label{section:surviving}}
\begin{figure*}
\centering
  \subfigure[2019Q1]{
    \includegraphics[width=0.39\textwidth]{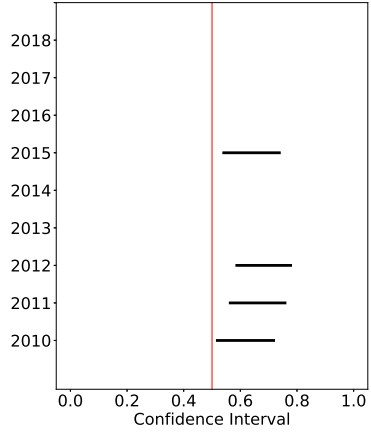}
  \label{figure:ci:final:PlC:2019Q1}}
  \subfigure[2019Q2]{
    \includegraphics[width=0.39\textwidth]{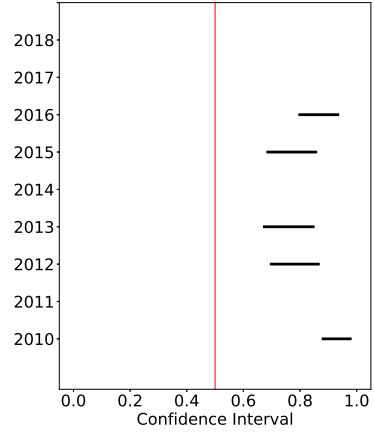}
  \label{figure:ci:final:PlC:2019Q2}}\\
  \subfigure[2019Q3]{
    \includegraphics[width=0.39\textwidth]{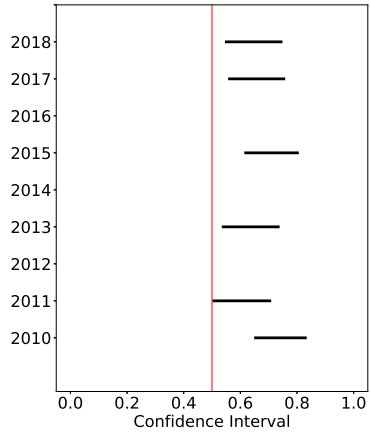}
  \label{figure:ci:final:PlC:2019Q3}}
  \subfigure[2019Q4]{
    \includegraphics[width=0.39\textwidth]{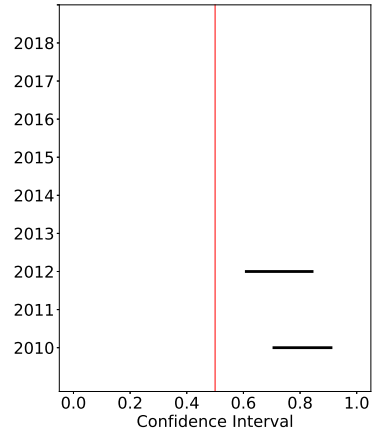}
  \label{figure:ci:final:PlC:2019Q4}}
\caption{Confidence intervals of $\pi_1$ for the pairwise comparisons with Binomial Sign Test for a Single Sample for $NO_2$ concentration at \textit{Plaza del Carmen} monitoring station. The non-significant test have been removed, as well as those pairwise comparisons with significant non-similar meteorology in the 2019 quarter and in the quarter from 2010 to 2018 under comparison.}
\label{figure:ci:final:PlC}
\end{figure*}

In Fig. \ref{figure:ci:final:PlC} the confidence intervals of $\pi_1$ for the pairwise comparisons with binomial sign test for a single sample for $NO_2$ concentration at Plaza del Carmen monitoring station under similar meteorological scenarios is presented. From Fig. \ref{figure:ci:binomial:PlC} to Fig. \ref{figure:ci:final:PlC}, the pairwise comparisons unfilling the following cuts have been removed:
\begin{itemize}
\item A significant reduction in the quarter of 2019 ---under the Binomial Sign Test for a Single Sample--- of $NO_2$ concentration.
\item A not significant more windy scenario in the quarter of 2019 ---under the Binomial Sign Test for a Single Sample---.
\item A not significant more rainy days scenario in the quarter of 2019 ---under the Chi-square Test for Homogeneity---.
\end{itemize}

As it can be observed, the number of surviving pairwise comparisons are 4 for the first quarter ---when only advertising was active---, 5 for the second quarter ---the first quarter with \textbf{Madrid Central} fully-active---, 6 in the third quarter ---mostly corresponding to summer period when $NO_2$ concentration is usually low---, and finally 2 for the fourth quarter ---corresponding with a quarter specially rainy---. 

Unfortunately, this procedure based on the significance of statistical tests does not provide a metric about the evolution of the distance of the quarters with the LEZ active and the previous quarters. This could be useful for evaluating the long-term evolution of the impact of the LEZ.

\subsection{GMM-based Analysis}

\begin{figure*}
{\renewcommand{\arraystretch}{1.0}
\rotatebox{90}{
\begin{minipage}[c][][c]{\textheight}
\centering
  \subfigure[2019Q1]{
    \includegraphics[width=0.425\textwidth]{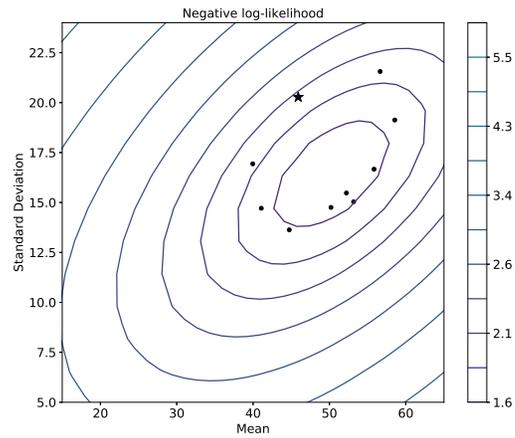}
  \label{figure:GMM:c1:2019Q1}}
  \subfigure[2019Q2]{
    \includegraphics[width=0.425\textwidth]{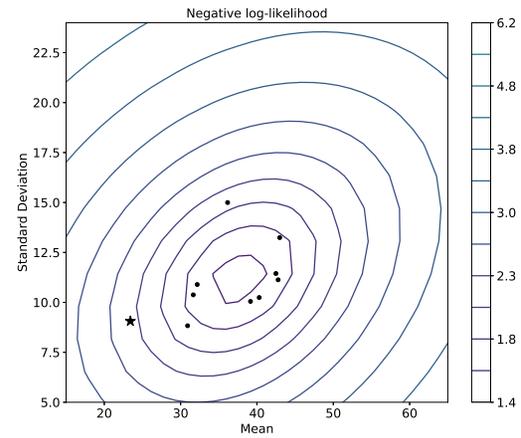}
  \label{figure:GMM:c1:2019Q2}}\\
  \subfigure[2019Q3]{
    \includegraphics[width=0.425\textwidth]{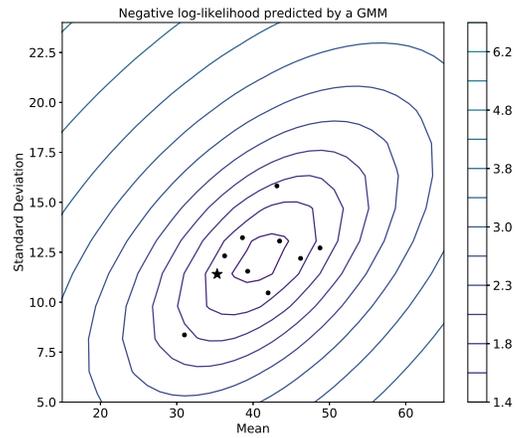}
  \label{figure:GMM:c1:2019Q3}}
  \subfigure[2019Q4]{
    \includegraphics[width=0.425\textwidth]{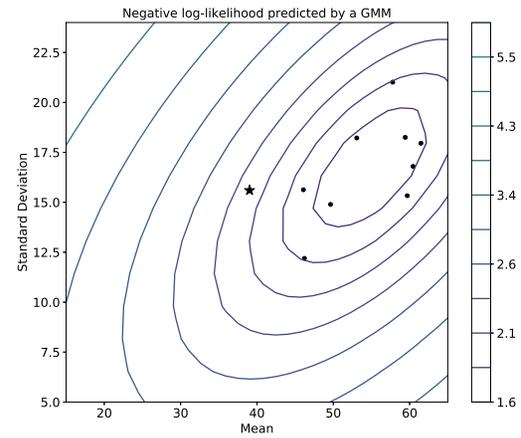}
  \label{figure:GMM:c1:2019Q4}}
\caption{GMM one component generated with the mean $\mu$ and standard deviation $\sigma$ of quarters from 2010 to 2018 (points). The contour lines of equal probability are also shown. The star corresponds to the $\mu$ and $\sigma$ of the quarters of 2019.}
\label{figure:GMM:c1}
\end{minipage}
}
}
\end{figure*}

GMM is able to build Gaussian probability distributions fitting the data. In this analysis the observations of each quarter are fitted to an univariante Gaussian probability distribution, from which the mean $\mu$ and standard deviation $\sigma$ are extracted. Then, these parameters are depicted in a plot and a bidimensional Gaussian probability distribution is built with $\mu$ and $\sigma$ of quarters from 2010 to 2018 (Fig. \ref{figure:GMM:c1}). The Gaussian probability distribution shown in this figure has been generated by fitting these points to a single probability distribution ---one component in GMM terminology---. Observations of 2019 quarters are also fitted to a Gaussian distribution, and its $\mu$ and $\sigma$ are also plotted (star point). The  contour lines of equal probability of the resulting distribution and the point corresponding to the $\mu$ and $\sigma$ 2019 quarter are shown. The relative position of this point in relation to the contour plots gives an intuition about the likelihood of this point to the probability distribution generated with the previous observations.

\subsubsection{Likelihood-Based Metric}
The described procedure allows numerically evaluating the likelihood of the point representing 2019 quarter observations to the probability distribution generated with the points representing the observations of the quarters from 2010 to 2018 (Table \ref{table:c1:loglikelihood}). 

As it can be observed in Fig. \ref{figure:GMM:c1:2019Q1}, 2019Q1 with \textbf{Madrid Central} still in advertising phase has a non-negligible probability to be part of the probability distribution coming from the first quarters of the periods 2010-2018. 

Oppositely, 2019Q2 has the second lowest likelihood to the probability distribution of points representing the observation from 2010 to 2018 (Fig. \ref{figure:GMM:c1:2019Q2}). This result is consistent with the reduction of $NO_2$ concentration observed in the Fig \ref{figure:ci:binomial:PlC:2019Q2}.

For the period 2019Q3, the logarithmic likelihood is the highest of the four periods analysed. With a high probability it can be stated that 2019Q3 belongs to the probability distribution representing the quarters of the previous years. This period corresponds with summer when sources of $NO_2$ are the lowest of the year and the removal by the high solar irradiation. Maybe this causes the lowest impact of the LEZ activation. 

Finally, the period 2019Q4, the logarithmic likelihood is the lowest one. This is consistent with the reduction of the concentration of $NO_2$ observed in 2019Q4 (Fig. \ref{figure:ci:binomial:PlC:2019Q4}).

\begin{table}
\caption{Logarithmic likelihood of 2019 quarters in the Gaussian probability distribution generated by GMM one component with the quarters of the years from 2010 to 2018.}
\label{table:c1:loglikelihood}
\centering 
\begin{tabular}{rr} \hline
Period & log(likelihood)  \\ \hline \hline
   Q1  & -7.52  \\ 
   Q2  & -8.44  \\ 
   Q3  & -4.59  \\ 
   Q4  & -9.01  \\  
\hline
\end{tabular}
\end{table}





\subsubsection{Jensen-Shannon Divergence-based Metric}
GMM is able to generate a bidimensional Gaussian probability distribution with the values of $\mu$ and $\sigma$ of the quarters from 2010 to 2018. Since GMM has been configured with one single component, its centroid: $\mu$ and $\sigma$, is representative of the quarters of the past years, from 2010 to 2018.  Therefore, the Jensen-Shannon divergence can be used to measure the distance between this centroid and the $\mu$ and $\sigma$ of the 2019 quarter.

In Table \ref{table:c1:DJS}, the Jensen-Shannon divergence between the $\mu$ and $\sigma$ of the 2019 quarters and the centroid of the quarters of the years from 2010 to 2018 is presented. Similarly to the likelihood-based analysis, 2019Q2 and 2019Q4 exhibit the largest differences with the previous periods. Again, the period previous to the full activation of \textbf{Madrid Central} ---2019Q1--- with only advertising activity does not show a clear impact in the reduction of the $NO_2$ concentration in \textit{Plaza del Carmen} monitoring station. And finally, the summer period of 2019 ---2019Q3--- with the lowest emissions exhibits a minor impact in the reduction of already low levels of $NO_2$ concentration. 

\begin{table}
\caption{Jensen-Shannon divergence for the $\mu$ and $\sigma$ of 2019 quarters and the $\mu$ and $\sigma$ of the centroids obtained with the GMM algorithm configured with 1 component.}
\label{table:c1:DJS}
\centering 
\begin{tabular}{rr} \hline
Period & Jensen-Shannon Divergence \\ \hline \hline
   Q1  & 0.07  \\ 
   Q2  & 1.06  \\ 
   Q3  & 0.12  \\ 
   Q4  & 0.48  \\ 
\hline
\end{tabular}
\end{table}

Likelihood and Jensen-Shannon divergence allow implementing metrics for evaluating how far the representation of a period ---$\mu$ and $\sigma$ of the fitting of the observation of the period to a Gaussian probability distribution--- is from the scenario previous to the LEZ activation. Thus, the impact of LEZ and its evolution along the years can be evaluated through this mechanism.

\subsubsection{Jensen-Shannon after Surviving $[NO_2]$ Reductions\label{section:JSDsurviving}}

In Table \ref{table:c2:DJSCuts}, the Jensen-Shannon Divergence between the $\mu$ and $\sigma$ of the 2019 quarters and the centroid of the quarters of the years from 2010 to 2018 is presented. Unlike the results shown in Table \ref{table:c1:DJS}, in Table  \ref{table:c2:DJSCuts} the coordinates of the GMM centroids have been calculated by using only the quarters for which similar meteorological conditions have been demonstrated (Fig. \ref{figure:ci:final:PlC}).

\begin{table}
\caption{Jensen-Shannon Divergence for the $\mu$ and $\sigma$ of 2019 quarters and the $\mu$ and $\sigma$ of the centroids obtained ith the GMM algorithm configured with one component, and using as input only the quarters with similar meteorological conditions to 2019.}\label{table:c2:DJSCuts}
\centering 
\begin{tabular}{rr} \hline
Period & Jensen-Shannon Divergence  \\ \hline \hline
   Q1  & 0.15  \\ 
   Q2  & 1.01  \\ 
   Q3  & 0.19  \\ 
   Q4  & 0.54  \\ \hline
\end{tabular}
\end{table}

The comparison between both tables shows changes in the distance, but keeping the order between the quarter of 2019 and the previous quarters. 
Since this distance has been calculated with quarters of similar meteorology, it is not biased by meteorological scenarios conducive to the elimination of pollution. 
This process can be repeated for the following years, and thus assess whether the initial impact of the LEZ is sustained over the years. 
Finally, the methodology presented is applicable to any pollutant and station, so potential negative impacts on the periphery of the LEZ can be evaluated.

\section{Results Around \textbf{Madrid Central} and Discussion \label{section:resultsOut}}

\begin{figure*}
\centering
    \includegraphics[width=1.\textwidth]{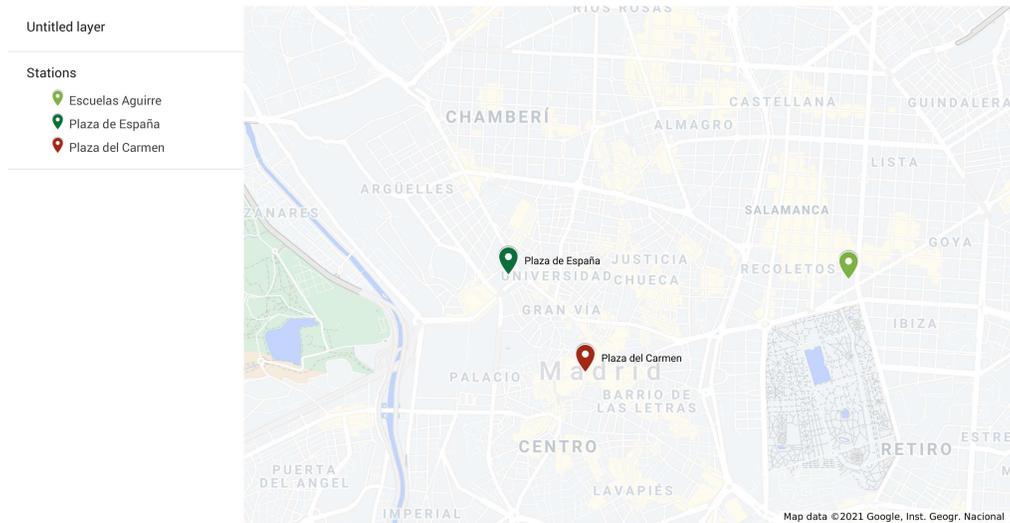}
\caption{Map of Madrid (Spain) with the positions of monitoring stations: \textit{Escuelas Aguirre}, \textit{Plaza del Carmen} and \textit{Plaza de Espa\~na}.}
\label{figure:map:3}
\end{figure*}

The analysis of the $[NO_2]$ and $[PM2.5]$ at \textit{Escuelas Aguirre} monitoring station, the $[NO_2]$ at \textit{Plaza de Espa\~na} around \textbf{Madrid Central} allows discerning the rise of negative effects due to an increment of the motor traffic around the LEZ (Fig. \ref{figure:map:3}). \textit{Escuelas Aguirre} is 2 km away from \textbf{Madrid Central}, and \textit{Plaza de Espa\~na} is only 1 km away. 
The periods evaluated include the three first quarters of the years from 2010 to 2019. Thus the three quarters where \textbf{Madrid Central} is fully operative, 2019Q2, 2019Q3 and 2019Q4, are included in the study. Furthermore, the period 2019Q1 is included as control.

\subsection{Escuelas Aguirre}

\begin{figure*}
{\renewcommand{\arraystretch}{1.0}
\rotatebox{90}{
\begin{minipage}[c][][c]{\textheight}
\centering
  \subfigure[2019Q1]{
    \includegraphics[width=0.47\textwidth]{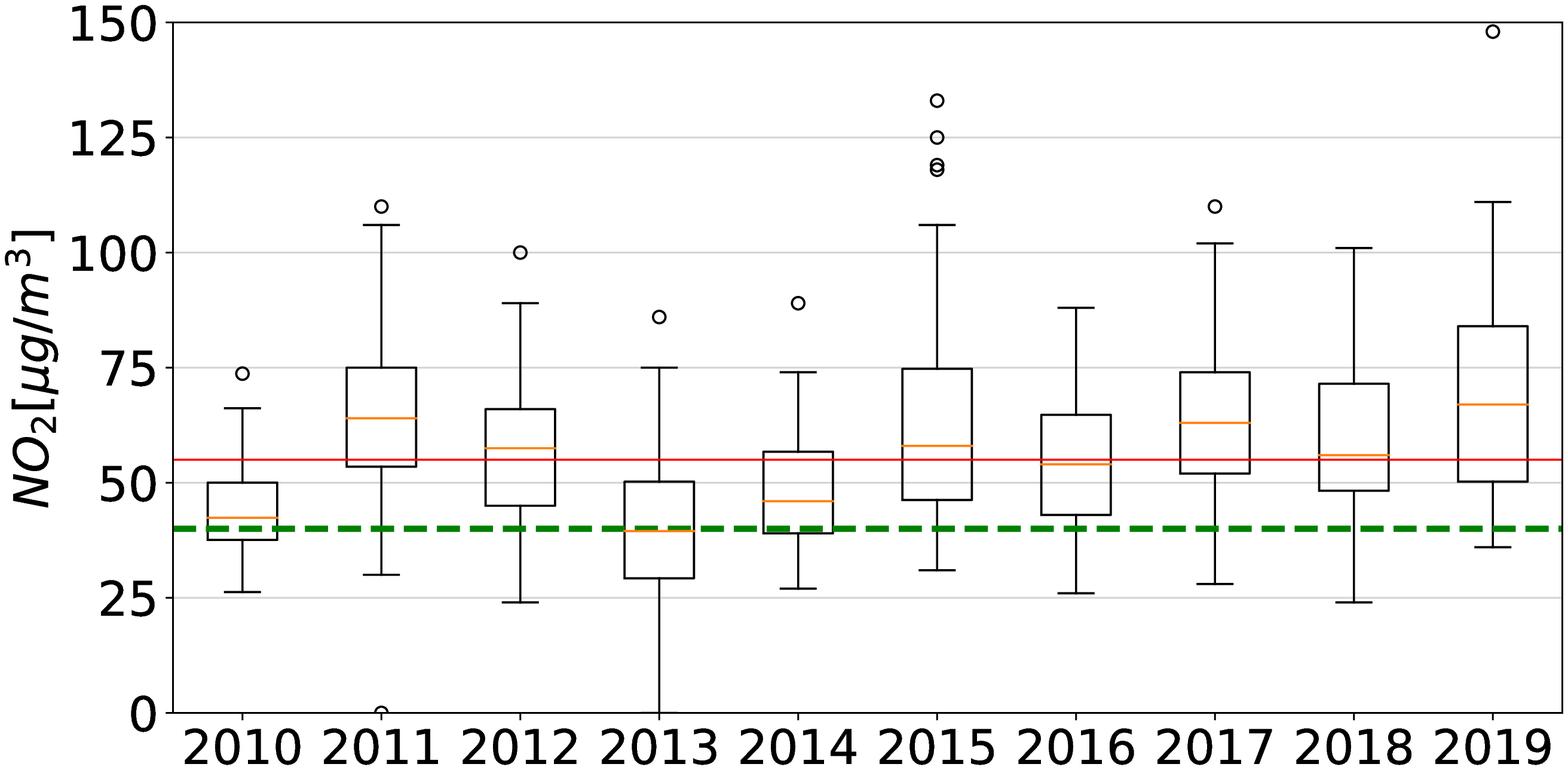}
  \label{figure:boxplot:2019Q1:NO2:EA}}
  \subfigure[2019Q2]{
    \includegraphics[width=0.47\textwidth]{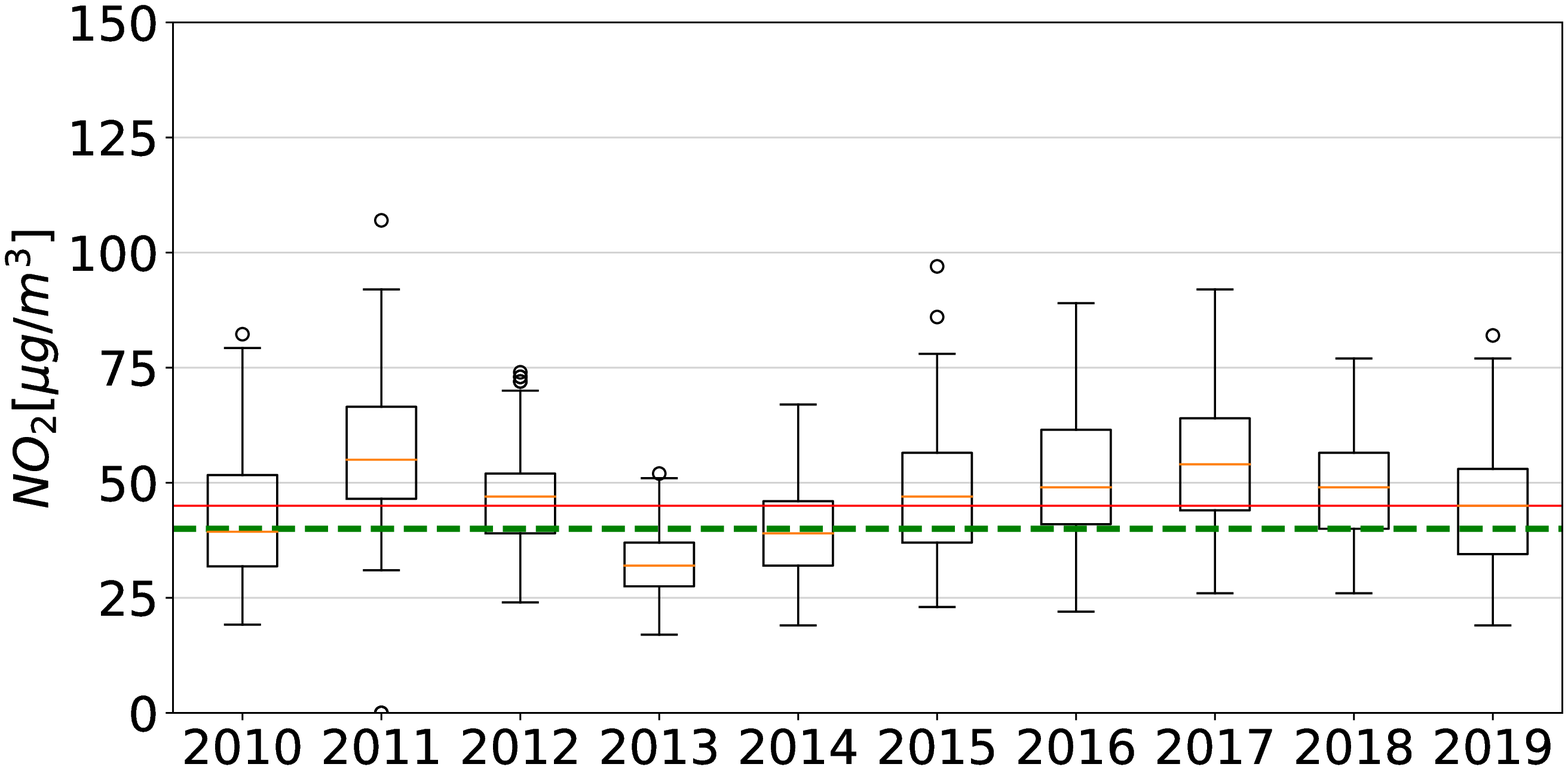}
  \label{figure:boxplot:2019Q2:NO2:EA}}
  \subfigure[2019Q3]{
    \includegraphics[width=0.47\textwidth]{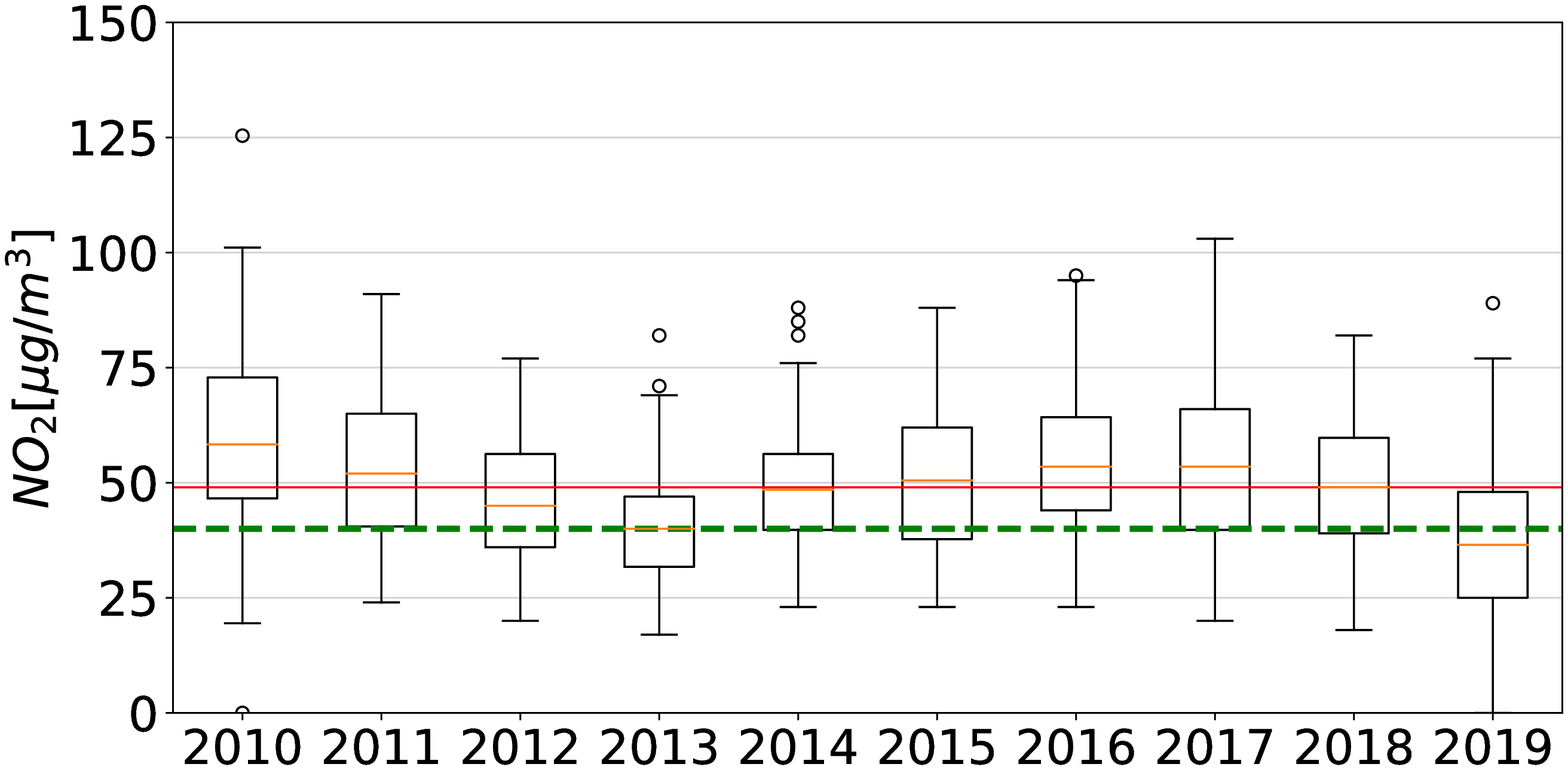}
  \label{figure:boxplot:2019Q3:NO2:EA}}
  \subfigure[2019Q4]{
    \includegraphics[width=0.47\textwidth]{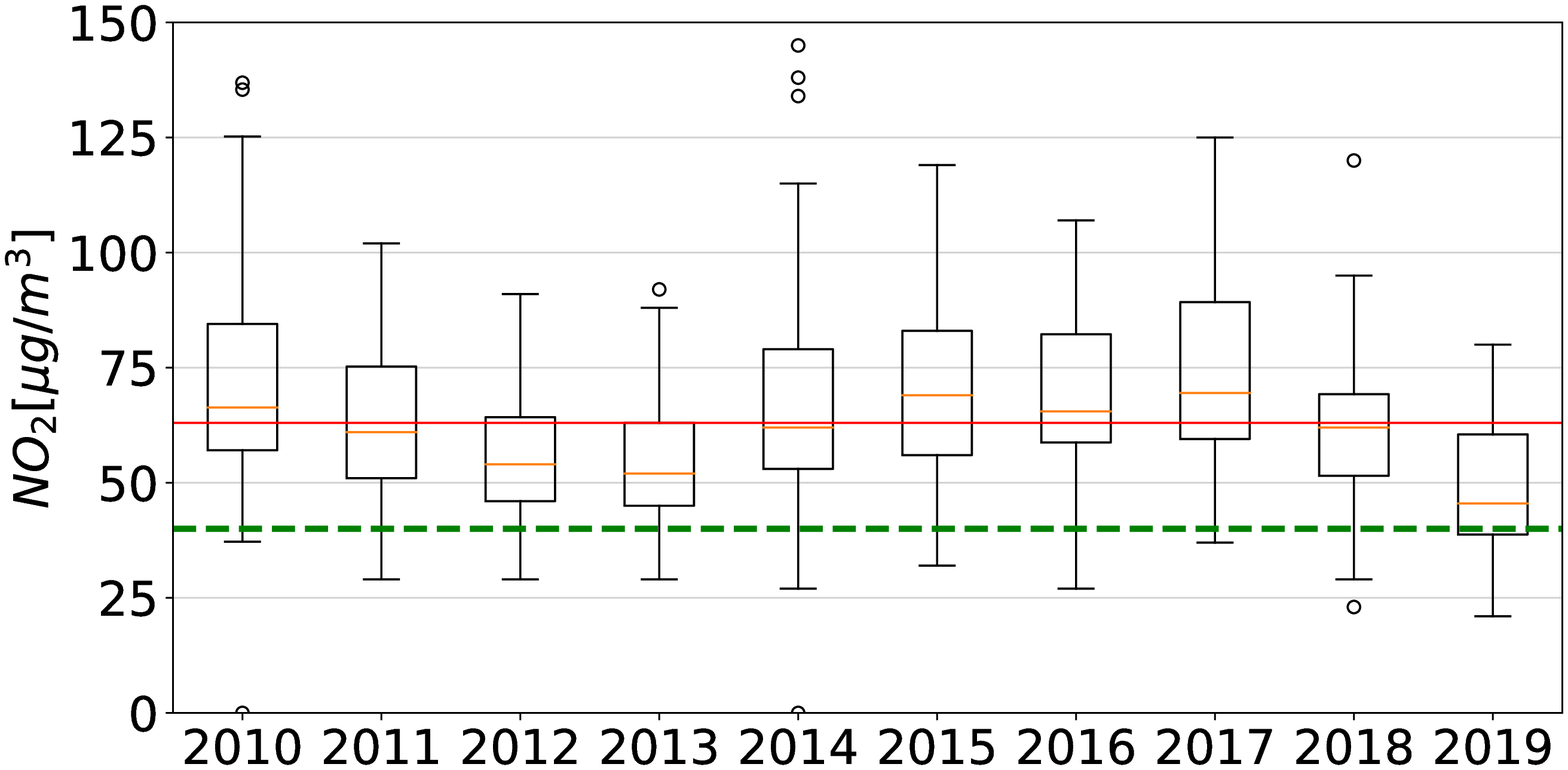}
  \label{figure:boxplot:2019Q4:NO2:EA}}
  \caption{Boxplot with the mean daily concentration of $NO_2$ for the periods 2019Q1 (Fig. \ref{figure:boxplot:2019Q1:NO2:EA}), 2019Q2 (Fig. \ref{figure:boxplot:2019Q2:NO2:EA}), 2019Q3 (Fig. \ref{figure:boxplot:2019Q3:NO2:EA}), and 2019Q4 (Fig. \ref{figure:boxplot:2019Q4:NO2:EA}) at \textit{Escuelas Aguirre} monitoring station. Red horizontal line shows the median of the values for the period 2010-2019.} 
\label{figure:boxplot:2019:NO2:EA}
\end{minipage}
        }
}
\end{figure*}

In Fig. \ref{figure:boxplot:2019:NO2:EA}, the daily mean $[NO_2]$ for the periods 2019Q1, 2019Q2, 2019Q3 and 2019Q4 at \textit{Escuelas Aguirre} station are shown. As it can be appreciated, $[NO_2]$ at \textit{Escuelas Aguirre} exhibits similarities to $[NO_2]$ at \textit{Plaza del Carmen} (Fig. \ref{figure:boxplot:2019:NO2:PlC}): no reduction in the first quarter (Fig. \ref{figure:boxplot:2019Q1:NO2:EA}), and reductions ---in some cases very clear (Fig. \ref{figure:boxplot:2019Q4:NO2:EA})--- in the remaining quarters that must be evaluated to ascertain their significance. 

\begin{figure*}
\centering
  \subfigure[2019Q1]{
    \includegraphics[width=0.39\textwidth]{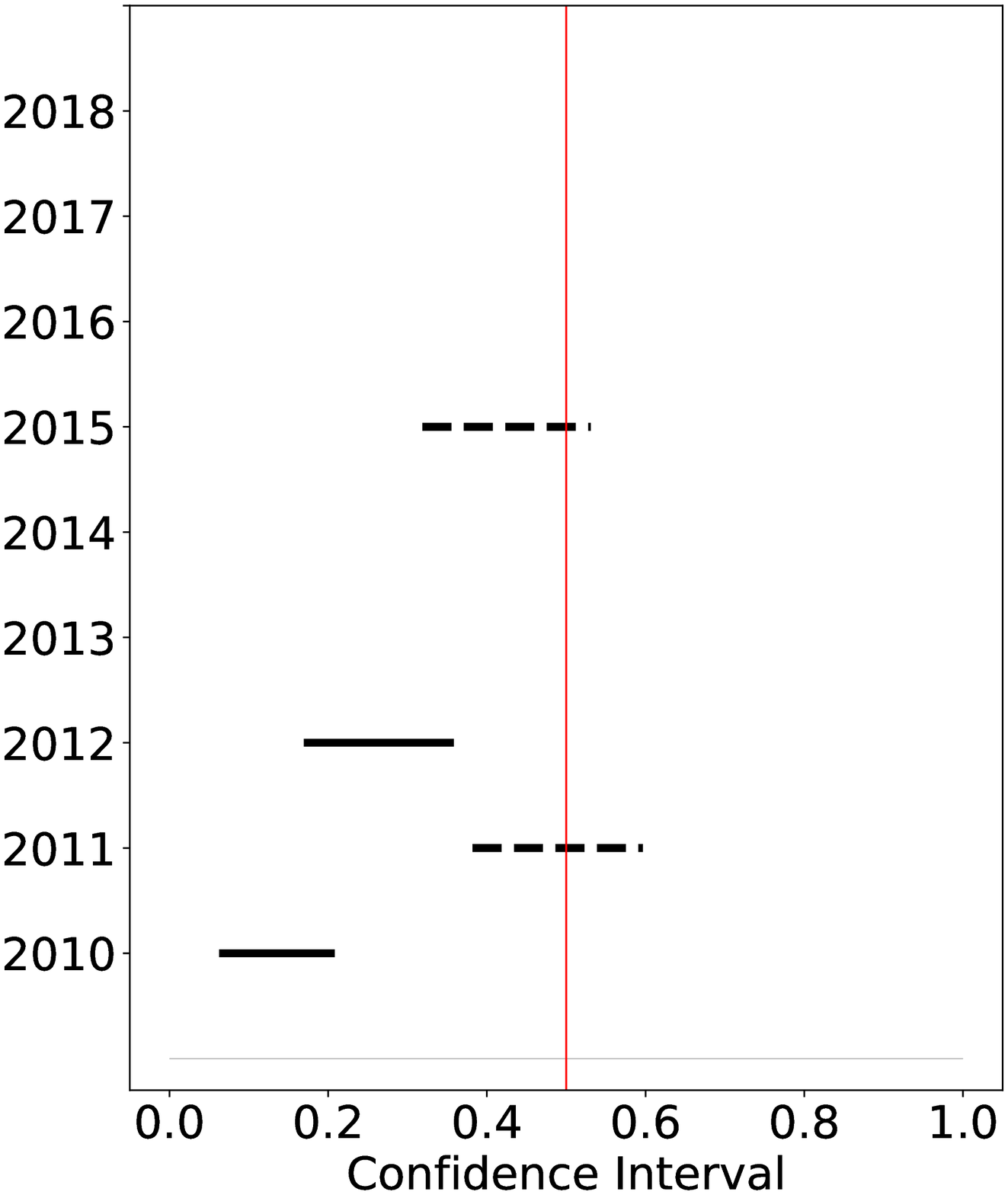}
  \label{figure:ci:final:EA:NO2:2019Q1}}
  \subfigure[2019Q2]{
    \includegraphics[width=0.39\textwidth]{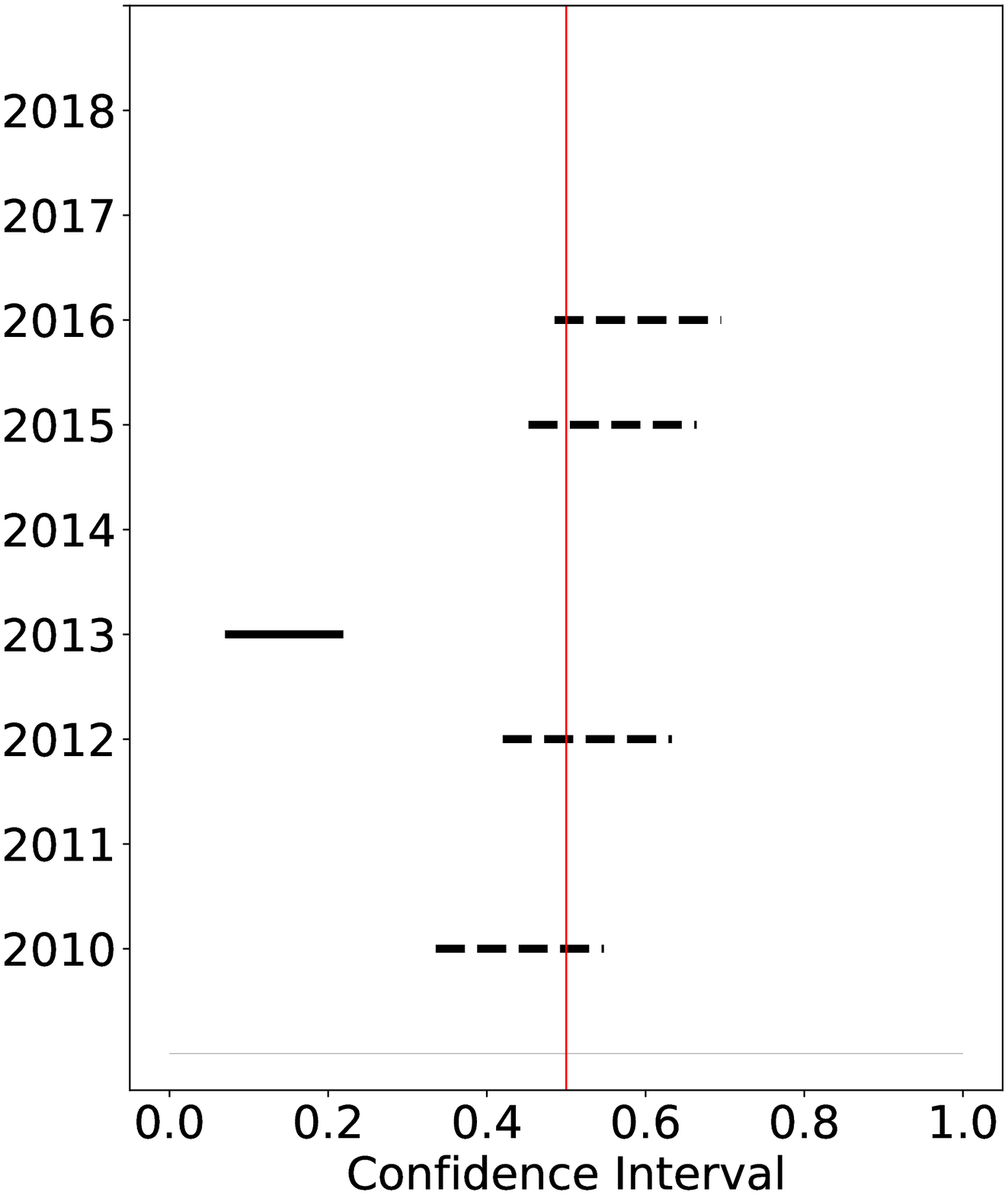}
  \label{figure:ci:final:EA:NO2:2019Q2}}\\
  \subfigure[2019Q3]{
    \includegraphics[width=0.39\textwidth]{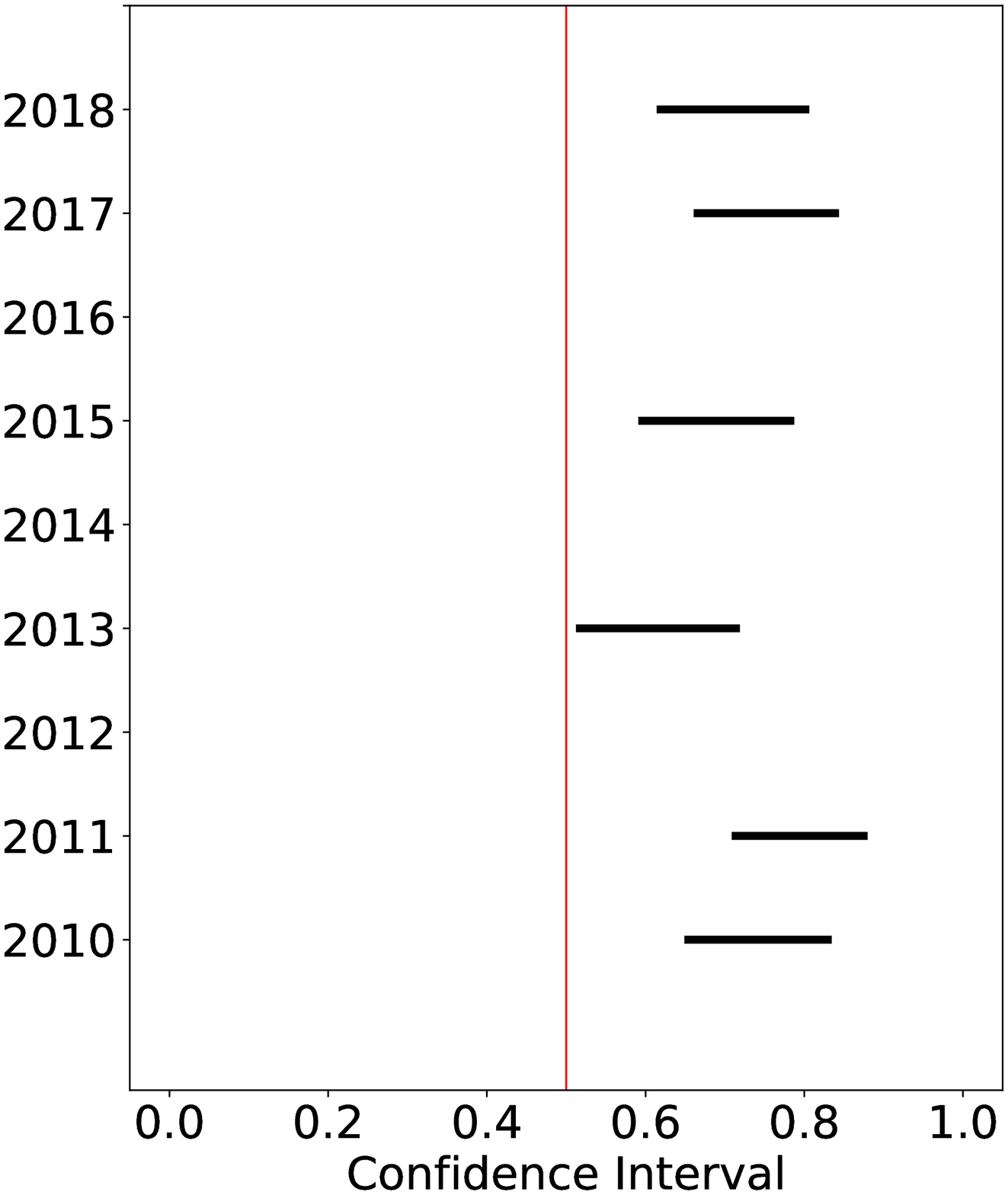}
  \label{figure:ci:final:EA:NO2:2019Q3}}
  \subfigure[2019Q4]{
    \includegraphics[width=0.39\textwidth]{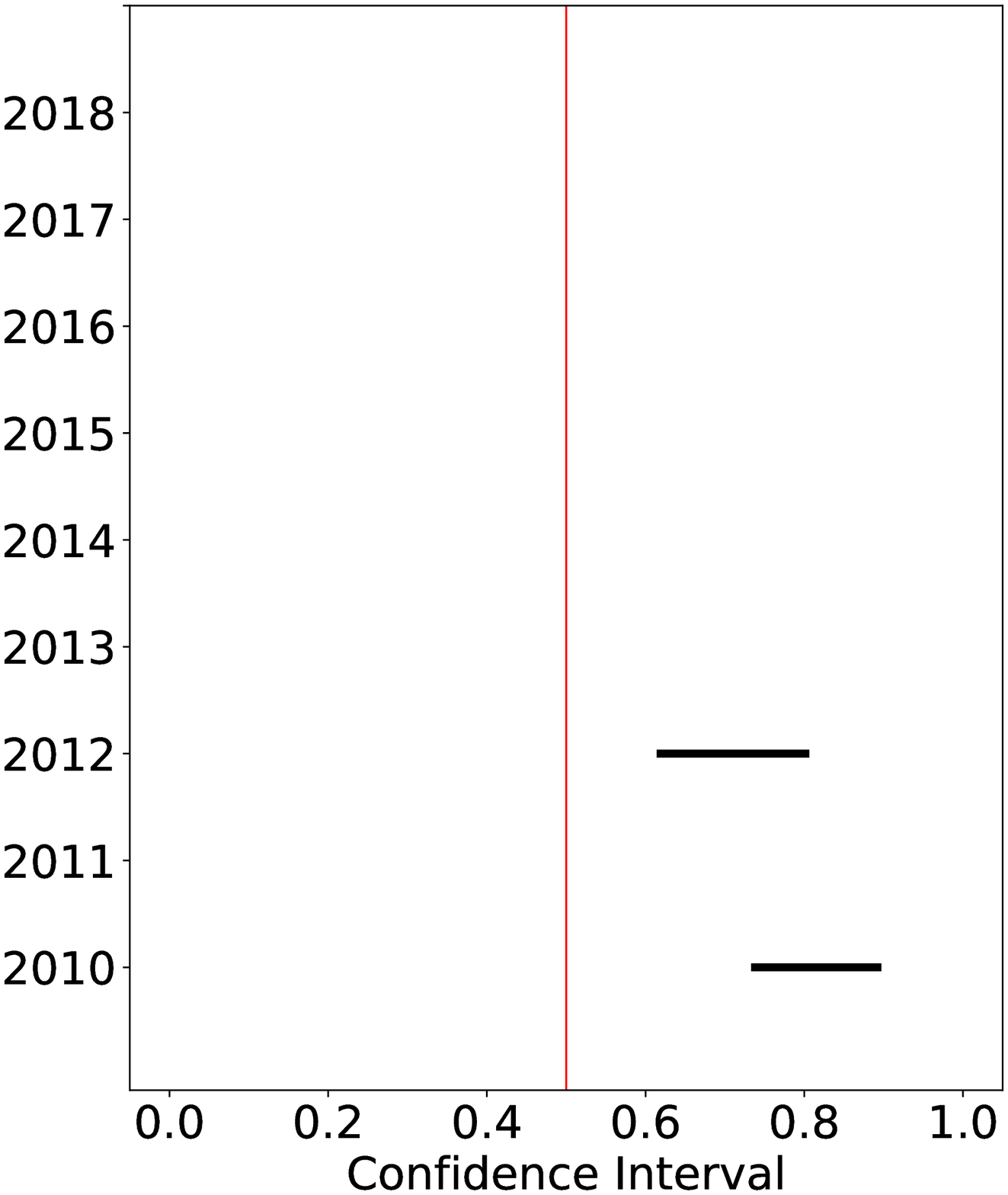}
  \label{figure:ci:final:EA:NO2:2019Q4}}
\caption{Confidence intervals of $\pi_1$ for the pairwise comparisons with Binomial Sign Test for a Single Sample for $NO_2$ concentration at \textit{Escuelas Aguirre} monitoring station. The non-significant test have been removed, as well as those pairwise comparisons with significant non-similar meteorology in the 2019 quarter and in the quarter from 2010 to 2018 under comparison.}
\label{figure:ci:final:EA:NO2}
\end{figure*}

In Fig. \ref{figure:ci:final:EA:NO2} the confidence intervals of $\pi_1$ for the pairwise comparisons with binomial sign test for a single sample for $[NO_2]$ concentration at \textit{Escuelas Aguirre} monitoring station under similar meteorological scenarios are presented. As it can be observed, the number of surviving pairwise comparisons are 4 for the first quarter ---when only advertising was active---, from which two pairwise comparisons show a significant increment of $[NO_2]$ (Fig. \ref{figure:ci:final:EA:NO2:2019Q1}). For the second quarter, they reduces to only one pairwise comparison (Fig. \ref{figure:ci:final:EA:NO2:2019Q2}). For the remaining quarter (Figs. \ref{figure:ci:final:EA:NO2:2019Q3} and \ref{figure:ci:final:EA:NO2:2019Q4}), significant increments in  the $[NO_2]$ are not longer appreciated. Oppositely, in the third and fourth quarters, in all the surviving pairwise comparisons significant reductions are observed. 

\begin{figure*}
{\renewcommand{\arraystretch}{1.0}
\rotatebox{90}{
\begin{minipage}[c][][c]{\textheight}
\centering
  \subfigure[2019Q1]{
    \includegraphics[width=0.47\textwidth]{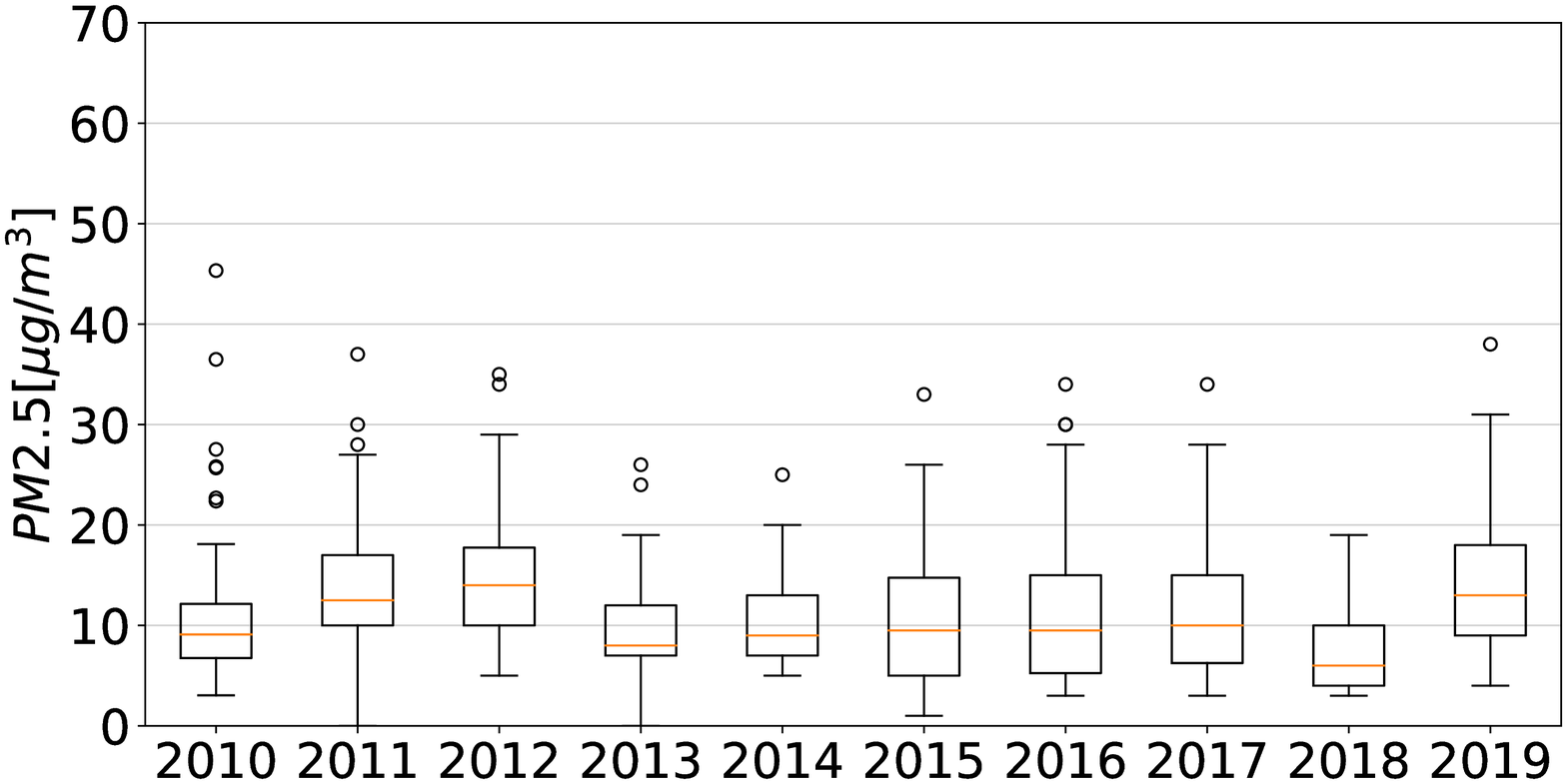}
  \label{figure:boxplot:2019Q1:PM25:EA}}
  \subfigure[2019Q2]{
    \includegraphics[width=0.47\textwidth]{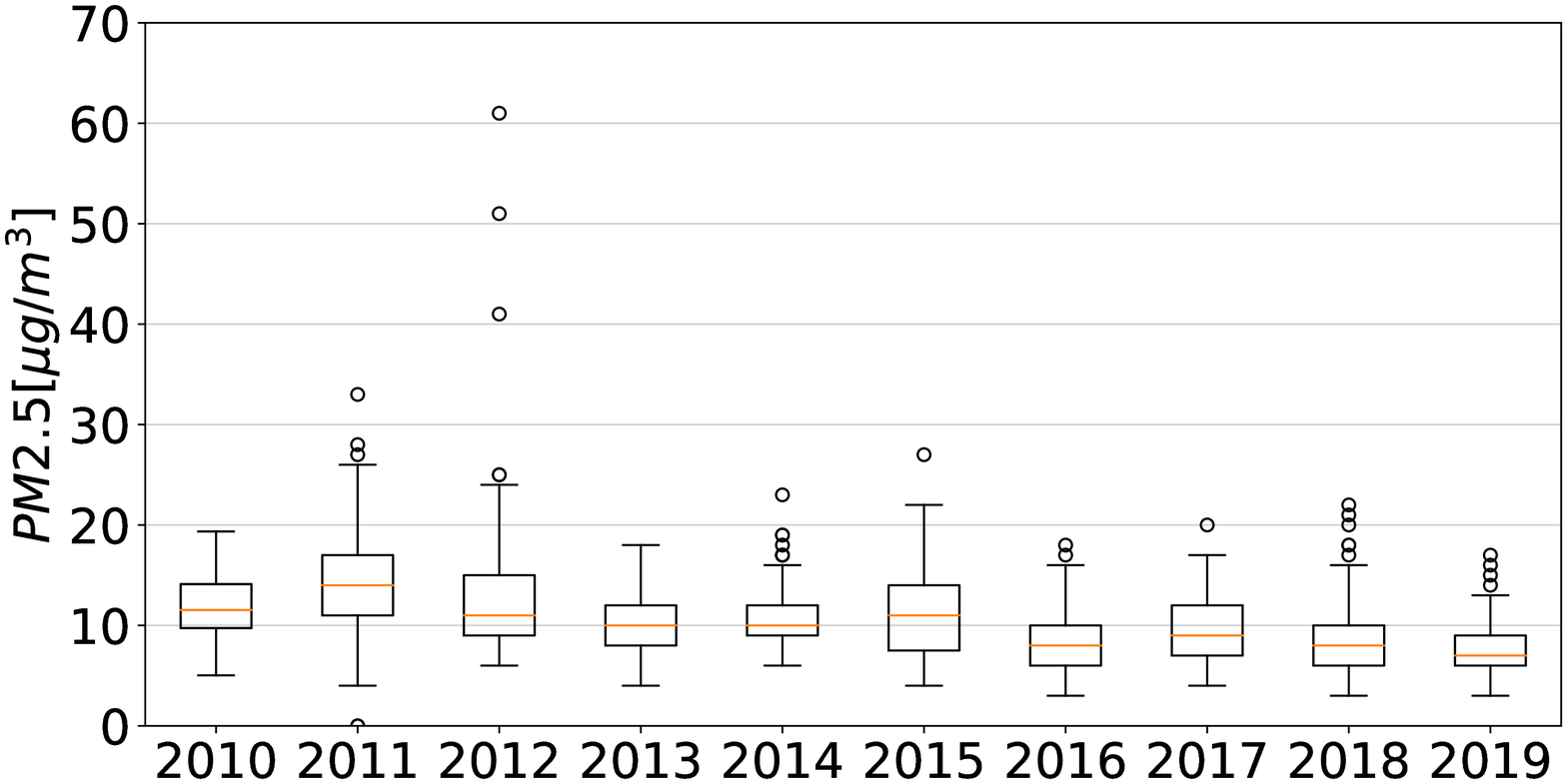}
  \label{figure:boxplot:2019Q2:PM25:EA}}
  \subfigure[2019Q3]{
    \includegraphics[width=0.47\textwidth]{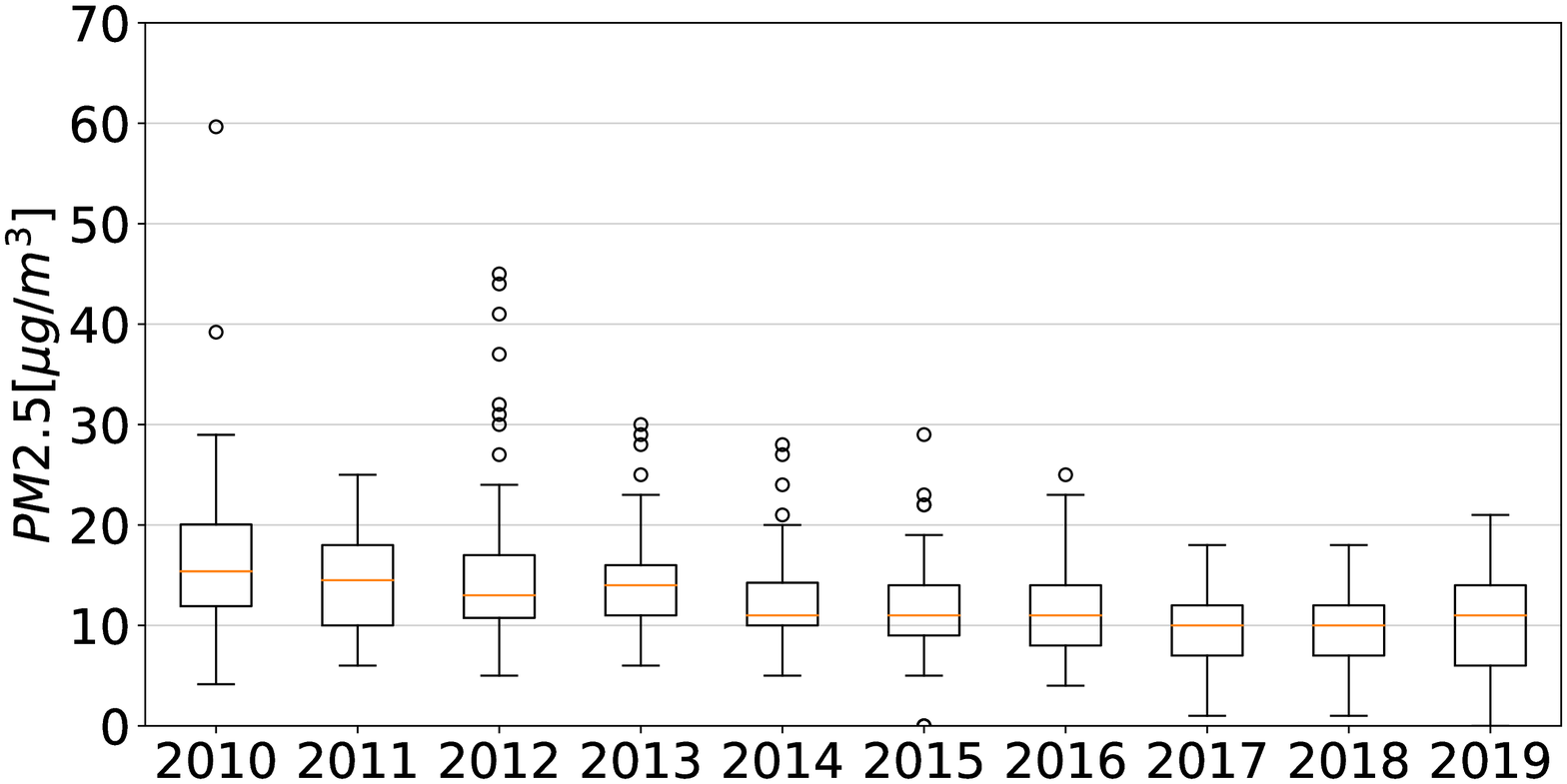}
  \label{figure:boxplot:2019Q3:PM25:EA}}
  \subfigure[2019Q4]{
    \includegraphics[width=0.47\textwidth]{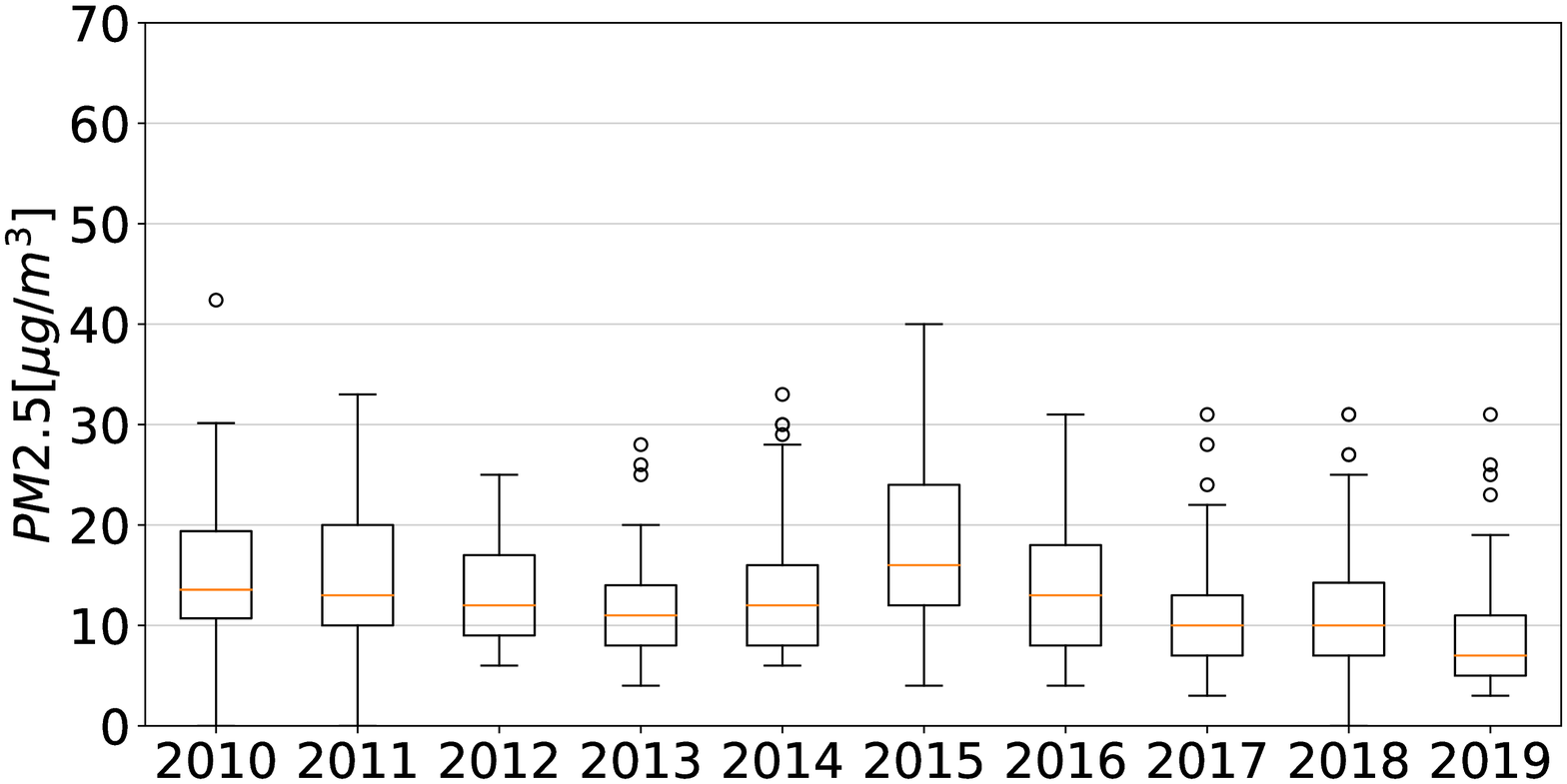}
  \label{figure:boxplot:2019Q4:PM25:EA}}
  \caption{Boxplot with the mean daily concentration of PM2.5 for the periods 2019Q1 (Fig. \ref{figure:boxplot:2019Q1:PM25:EA}), 2019Q2 (Fig. \ref{figure:boxplot:2019Q2:PM25:EA}), 2019Q3 (Fig. \ref{figure:boxplot:2019Q3:PM25:EA}), and 2019Q4 (Fig. \ref{figure:boxplot:2019Q4:PM25:EA}) at \textit{Escuelas Aguirre} monitoring station. Red horizontal line shows the median of the values for the period 2010-2019.} 
\label{figure:boxplot:2019:PM25:EA}
\end{minipage}
        }
}
\end{figure*}

In Fig. \ref{figure:boxplot:2019:PM25:EA}, the daily mean $[PM2.5]$ for the periods 2019Q1, 2019Q2, 2019Q3 and 2019Q4 at \textit{Escuelas Aguirre} are shown. As can be observed, measurements for the quarters of 2019 are similar to previous years, with slight higher values for the observations of period 2019Q1 (Fig. \ref{figure:boxplot:2019Q1:PM25:EA}), and a smooth trend to lower values for the periods 2019Q2 (Fig. \ref{figure:boxplot:2019Q2:PM25:EA}), 2019Q3 (Fig. \ref{figure:boxplot:2019Q3:PM25:EA}), and 2019Q4 (Fig. \ref{figure:boxplot:2019Q4:PM25:EA}). 

In \textit{Escuelas Aguirre} $[PM2.5]$ are also avaible for performing additional comparisons. Both pollutants, $[PM2.5]$ and $[NO_2]$, can be associated to motor traffic, although not exclusively. As it can observed, in the first quarter two pairwise comparisons show significant increments in the pollutant concentration, whereas two other comparisons are not significant. In the remaining quarters, in most of the cases the comparisons exhibit significant reductions of the pollutant. As a consequence of the analysis of $[PM2.5]$ and $[NO_2]$ in  \textit{Escuelas Aguirre}, the statement of degradation of the air quality at this place as results of the activation of the LEZ is not supported by the analyses. 


\begin{figure*}
\centering
  \subfigure[2019Q1]{
    \includegraphics[width=0.39\textwidth]{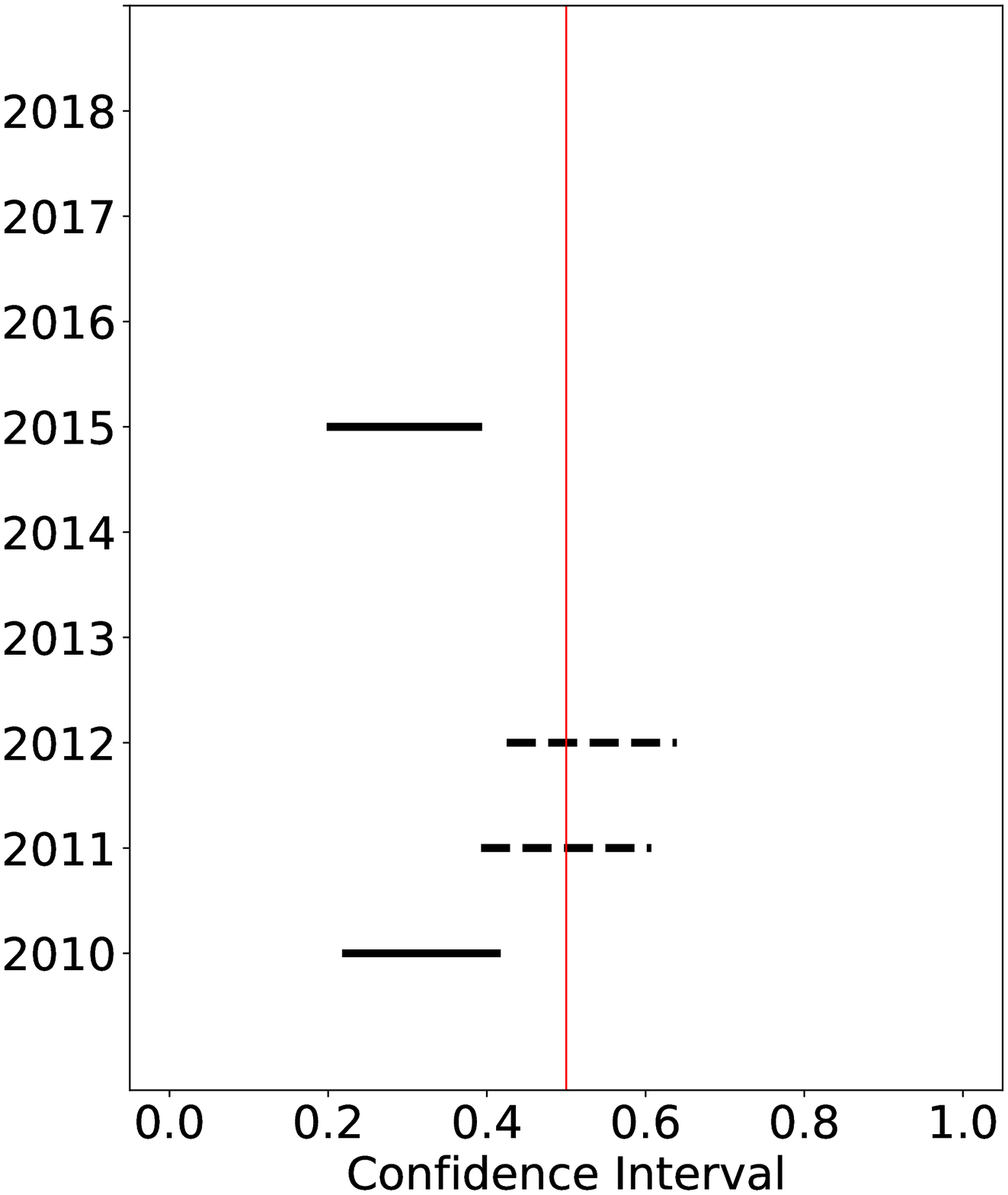}
  \label{figure:ci:final:EA:NO2:2019Q1}}
  \subfigure[2019Q2]{
    \includegraphics[width=0.39\textwidth]{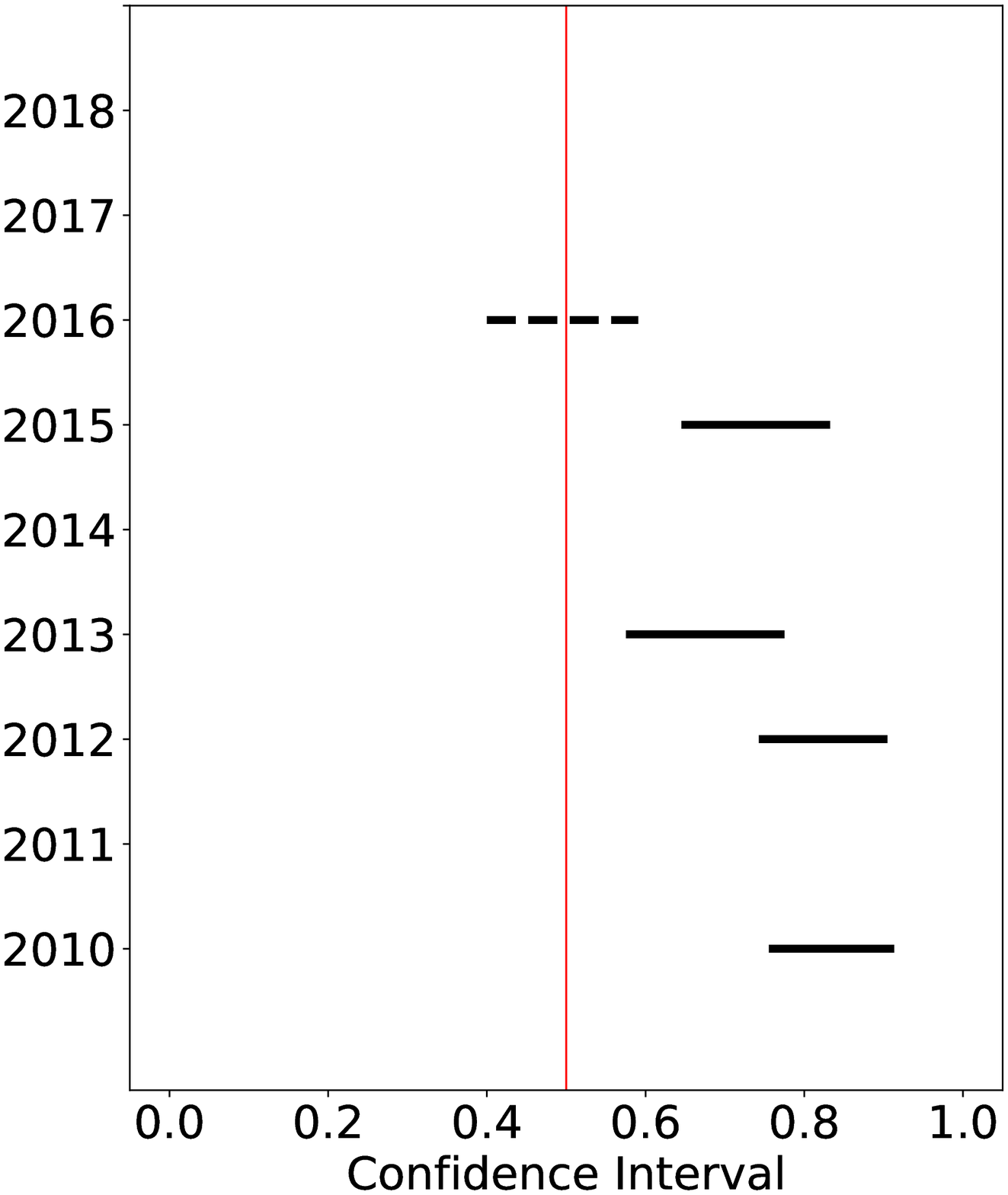}
  \label{figure:ci:final:EA:NO2:2019Q2}}\\
  \subfigure[2019Q3]{
    \includegraphics[width=0.39\textwidth]{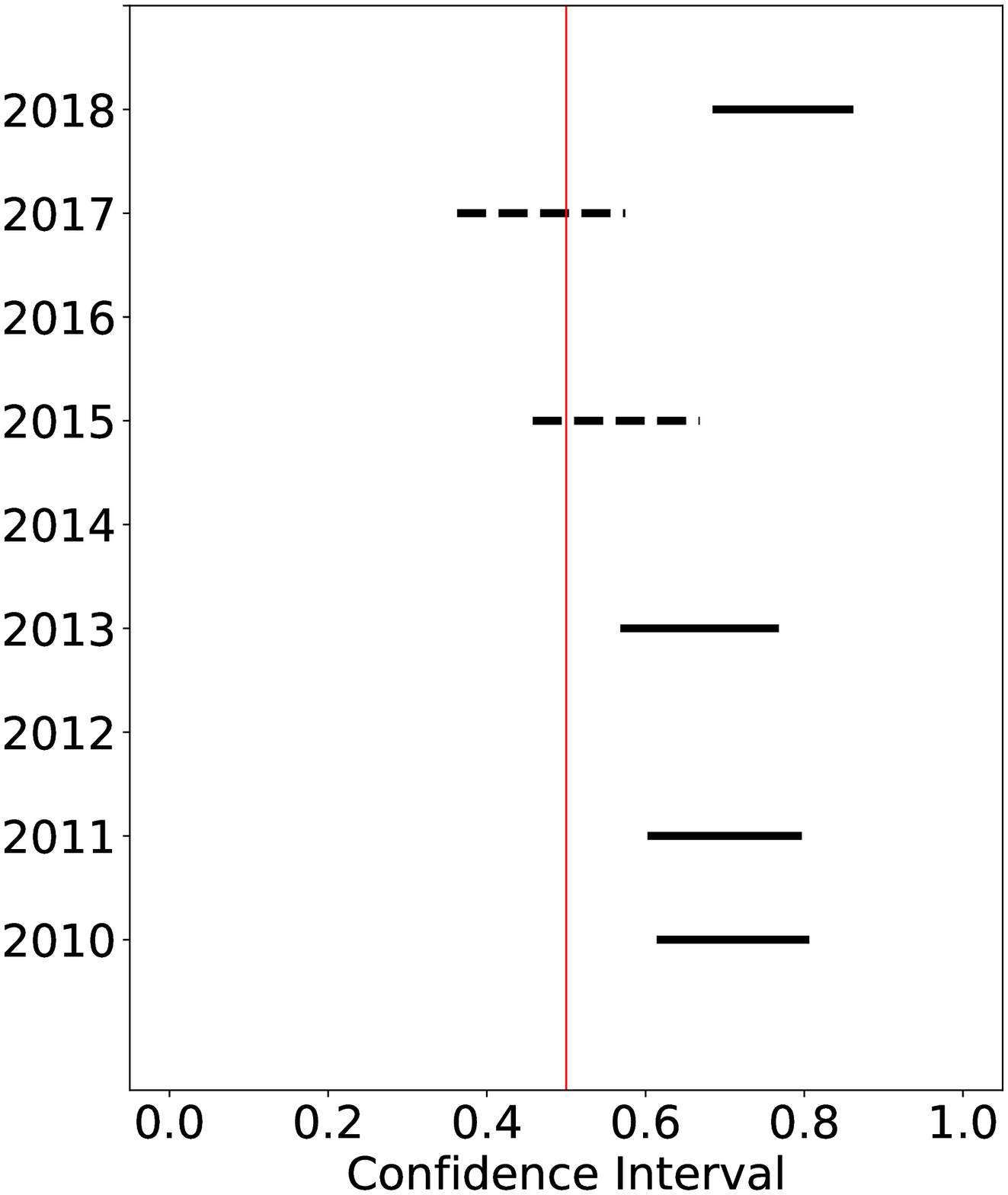}
  \label{figure:ci:final:EA:NO2:2019Q3}}
  \subfigure[2019Q4]{
    \includegraphics[width=0.39\textwidth]{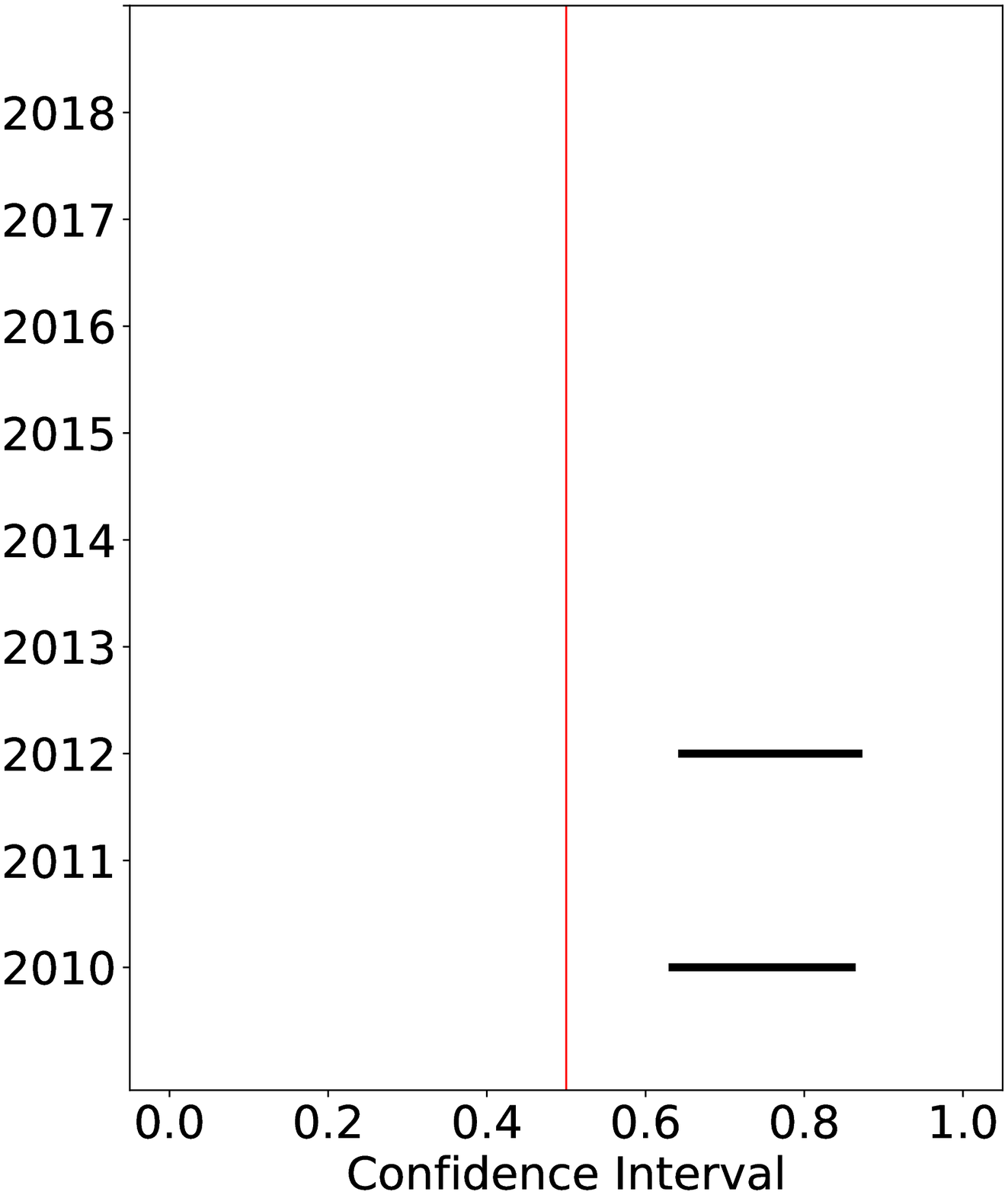}
  \label{figure:ci:final:EA:NO2:2019Q4}}
\caption{Confidence intervals of $\pi_1$ for the pairwise comparisons with Binomial Sign Test for a Single Sample for $PM2.5$ concentration at \textit{Escuelas Aguirre} monitoring station. The non-significant test have been removed, as well as those pairwise comparisons with significant non-similar meteorology in the 2019 quarter and in the quarter from 2010 to 2018 under comparison.}
\label{figure:ci:final:EA:PM2.5}
\end{figure*}

In Fig. \ref{figure:ci:final:EA:PM2.5} the confidence intervals of $\pi_1$ for the pairwise comparisons with binomial sign test for a single sample for $[PM2.5]$ concentration at \textit{Escuelas Aguirre} monitoring station under similar meteorological scenarios are presented. 
As it can be observed, 4 pairwise comparison survives to the comparison under similar meteorological scenarios. Two of them indicate a significant degradation of the air quality (Fig. \ref{figure:ci:final:EA:NO2:2019Q1}). In the second and the third quarters (Figs. \ref{figure:ci:final:EA:NO2:2019Q2} and \ref{figure:ci:final:EA:NO2:2019Q3}), four pairwise comparisons exhibit a significant reduction of $[PM2.5]$ after the activation of the LEZ. In the fourth quarter (Fig. \ref{figure:ci:final:EA:NO2:2019Q4}), the two only surviving comparisons show a significant improvement of the air quality.

\subsection{Plaza de Espa\~na}

\begin{figure*}
{\renewcommand{\arraystretch}{1.0}
\rotatebox{90}{
\begin{minipage}[c][][c]{\textheight}
\centering
  \subfigure[2019Q1]{
    \includegraphics[width=0.47\textwidth]{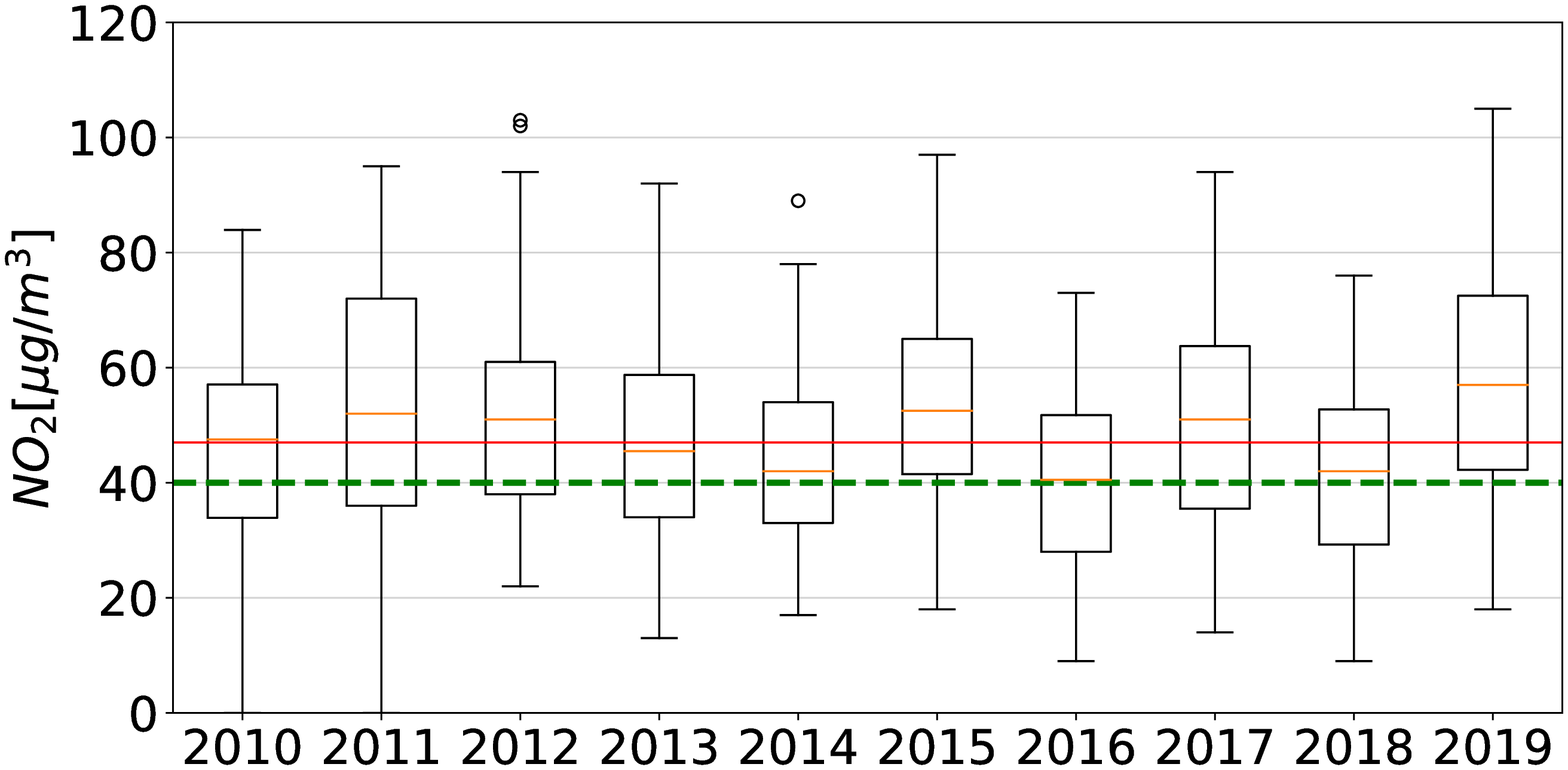}
  \label{figure:boxplot:2019Q1:NO2:PlE}}
  \subfigure[2019Q2]{
    \includegraphics[width=0.47\textwidth]{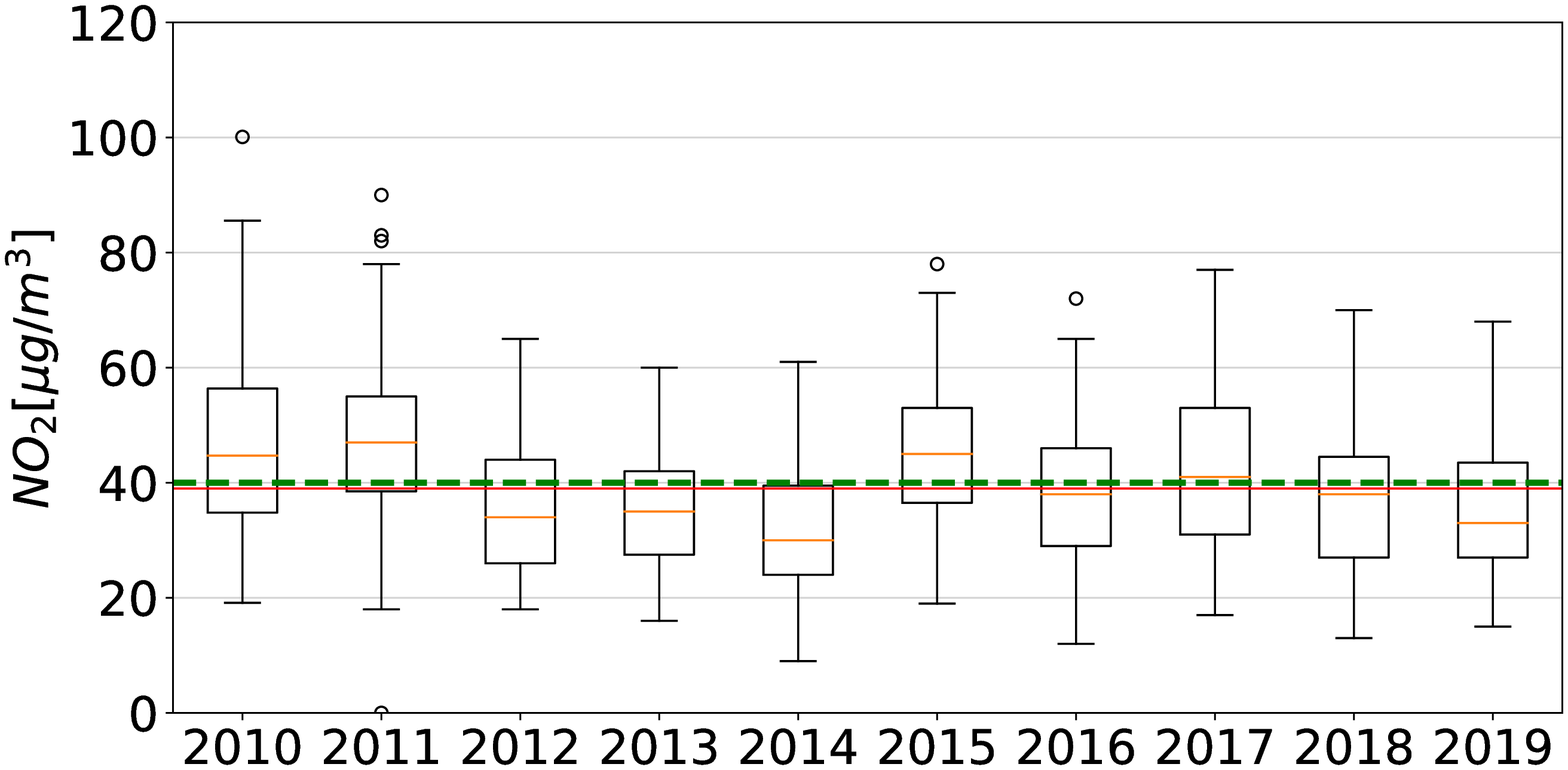}
  \label{figure:boxplot:2019Q2:NO2:PlE}}
  \subfigure[2019Q3]{
    \includegraphics[width=0.47\textwidth]{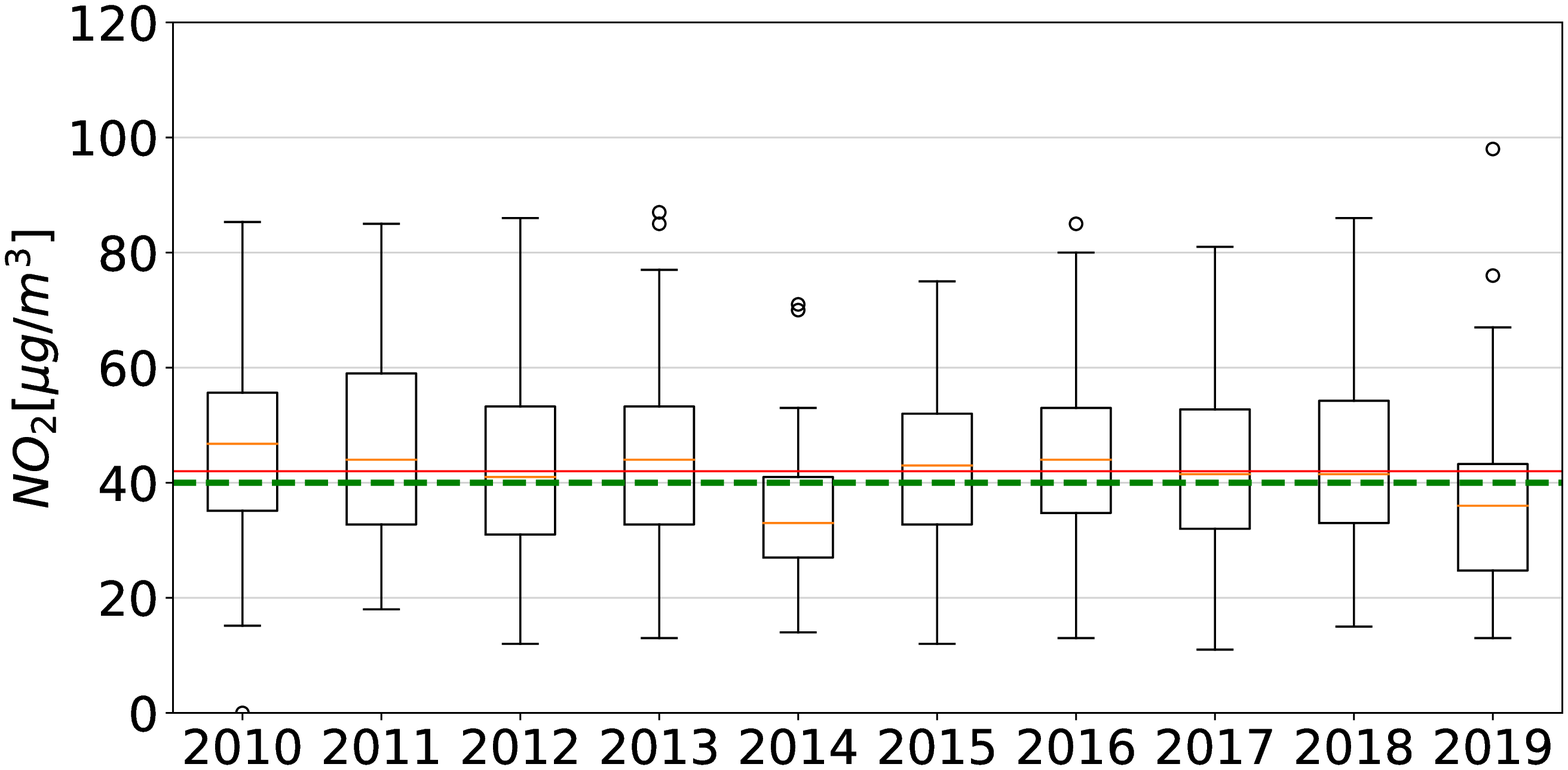}
  \label{figure:boxplot:2019Q3:NO2:PlE}}
  \subfigure[2019Q4]{
    \includegraphics[width=0.47\textwidth]{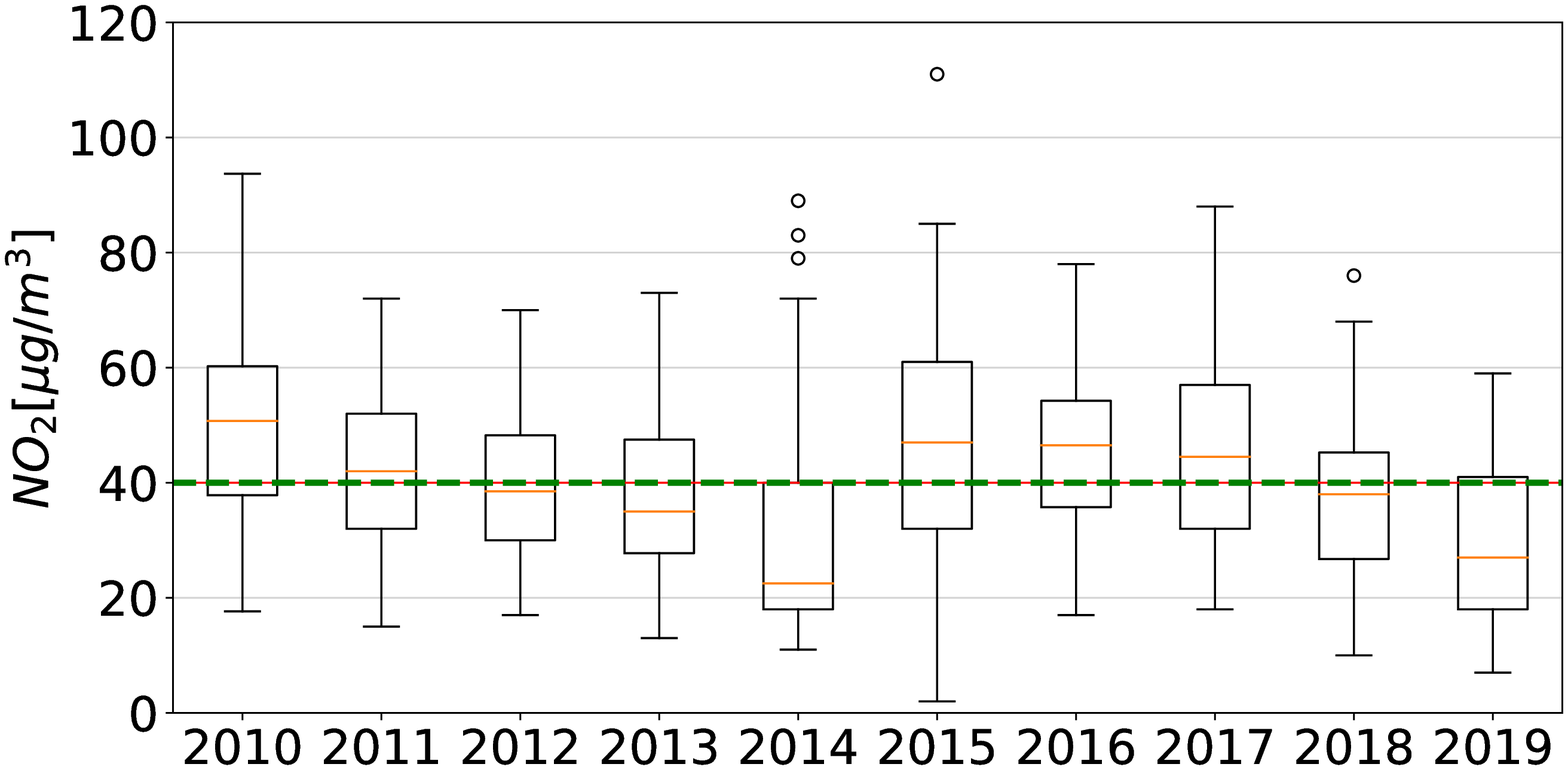}
  \label{figure:boxplot:2019Q4:NO2:PlE}}
  \caption{Boxplot with the mean daily concentration of $NO_2$ for the periods 2019Q1 (Fig. \ref{figure:boxplot:2019Q1:NO2:PlE}), 2019Q2 (Fig. \ref{figure:boxplot:2019Q2:NO2:PlE}), 2019Q3 (Fig. \ref{figure:boxplot:2019Q3:NO2:PlE}), and 2019Q4 (Fig. \ref{figure:boxplot:2019Q4:NO2:PlE}) at \textit{Plaza de Espa\~na} monitoring station. Red horizontal line shows the median of the values for the period 2010-2019.} 
\label{figure:boxplot:2019:NO2:PlE}
\end{minipage}
        }
}
\end{figure*}

In Fig. \ref{figure:boxplot:2019:NO2:PlE}, the daily mean $[NO_2]$ for the periods 2019Q1, 2019Q2, 2019Q3 and 2019Q4 at \textit{Plaza de Espa\~na} are shown. Again, similarities appear in the pattern of $[NO_2]$ at this placement and inside the LEZ: first quarter does not show any reduction (Fig. \ref{figure:boxplot:2019Q1:NO2:PlE}), whereas for the other quarters, some reductions are observed. 

\begin{figure*}
\centering
  \subfigure[2019Q1]{
    \includegraphics[width=0.39\textwidth]{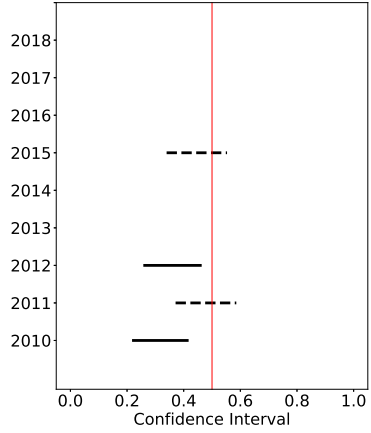}
  \label{figure:ci:final:PlE:2019Q1}}
  \subfigure[2019Q2]{
    \includegraphics[width=0.39\textwidth]{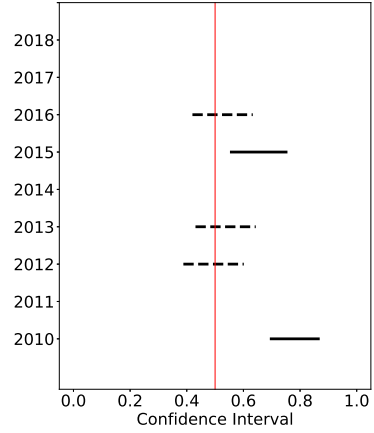}
  \label{figure:ci:final:PlE:2019Q2}}\\
  \subfigure[2019Q3]{
    \includegraphics[width=0.39\textwidth]{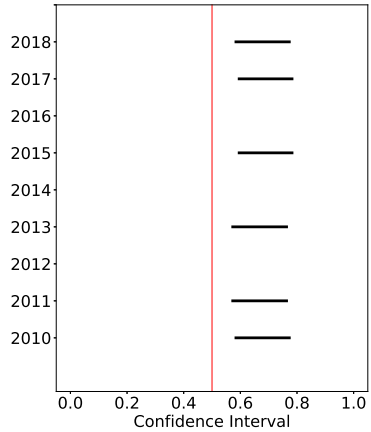}
  \label{figure:ci:final:PlE:2019Q3}}
  \subfigure[2019Q4]{
    \includegraphics[width=0.39\textwidth]{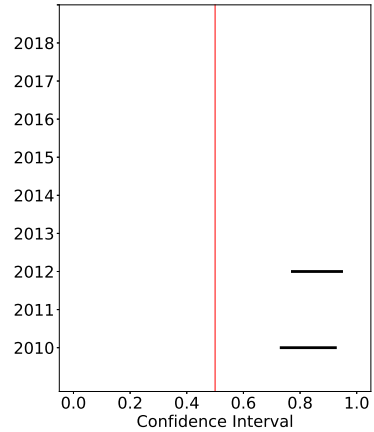}
  \label{figure:ci:final:PlE:2019Q4}}
\caption{Confidence intervals of $\pi_1$ for the pairwise comparisons with Binomial Sign Test for a Single Sample for $NO_2$ concentration at \textit{Plaza de Espa\~na} monitoring station. The non-significant test have been removed, as well as those pairwise comparisons with significant non-similar meteorology in the 2019 quarter and in the quarter from 2010 to 2018 under comparison.}
\label{figure:ci:final:PlE:NO2}
\end{figure*}

In Fig. \ref{figure:ci:final:PlE:NO2} the confidence intervals of $\pi_1$ for the pairwise comparisons with binomial sign test for a single sample for $[PM2.5]$ concentration at \textit{Plaza de Espa\~na} monitoring station under similar meteorological scenarios are presented. As it can be observed, for the first quarte two pairwise comparisons ---from the four surviving under similar meteorological scenarios--- exhibit a significant increment of $[NO_2]$. For the rest of the quartes, not more cases of significant increment of this pollutant are found. Therefore, any statement involving a degradation of the air quality at \textit{Plaza de Espa\~na} as a consequence of the implementation of the LEZ is not supported by the analysis.


\section{Conclusions}\label{section:conclusions}

In this paper, efforts for evaluating the impact of the activation of \textbf{Madrid Central} are presented. Today citizens are concerned by the impact of the air-quality in urban areas and their health. And, for this reason, they demand actions mitigating the high level of the most dangerous pollutants. LEZ are being implanted in the city center for reducing the concentration of these pollutants. At the end of 2018, Madrid established a restricted traffic area of 472 hectares.


After the first year of operation, it is the right time to assess its impact. For this, in this work two methodologies are proposed. The first one is based on statistical tests such as the Binomial Sign Test for a Single Sample and the Chi-square Test for Homogeneity. And, the second one applies Gaussian Mixture Models on the representative parameters of the past periods and the Jensen-Shannon divergence. Finally, both methods are merged for producing an unbiased metric.

In comparison with the previous observations in the period 2010-2018, both the metric shows relevant reductions on the $[NO_2]$ in the second and fourth quarters of 2019. Unfortunately, the large number of rainy days in 2019Q4 removes many pairwise comparisons due to not similar meteorological conditions, so the comparison is undertaken only with two quarters.

Besides, minor reductions can be claimed in the first and third  quarters. During the first quarter of 2019 \textbf{Madrid Central} is not fully operational: infringements are reported but not fined. Whereas the third quarter, corresponding to the summer period, where the sources of $NO_2$ are the lowest, and therefore, critical reductions can not be expected. From the analyses, it can be stated the positive reduction of the concentration of $[NO_2]$ during the first year of \textbf{Madrid Central}. 

Furthermore, the application of the proposed methodology to two other monitoring stations around \textbf{Madrid Central} allows discarding any increment in the $[NO_2]$ and in the $[PM2.5]$ due to the activation of the LEZ.

The proposed methodology has limitations. One of these limitations is the inability to separate the influence of various factors on the variation of the concentration of a pollutant: for example, activation of LEZ and favourable weather. 
Currently the quarter of the previous years is maintained for the pairwise comparison if the meteorological conditions are similar, otherwise it is removed.  
To have the capability to separate the impact of LEZ activation and a meteorological scenario: favourable or unfavourable for the reduction of a pollutant is proposed as future work.


\section*{Acknowledgements}
The research leading to these results has received funding by the Spanish Ministry of Economy and Competitiveness (MINECO) for funding support through the grant MDM-2015-0509: "Unidad de Excelencia Mar\'ia de Maez\-tu": CIEMAT - F\'ISICA DE PART\'ICULAS. 

Author gratefully acknowledges Dr. Fernando Mart\'in Llorente for his valuable suggestions and
discussions.

\end{document}